\documentclass[preprint]{aastex}
\usepackage{graphicx}
\usepackage{subfigure}
\usepackage{txfonts}
\usepackage{natbib}
\usepackage{longtable}
\usepackage{hyperref}

\newcommand{\degree}{\ensuremath{^\circ}}
\newcommand{\HII}{\ion{H}{2}}
 
\bibpunct{(}{)}{;}{a}{}{,}
\begin{document}

\title{Cold Dust in Hot Regions}

\author{Gopika Sreenilayam\altaffilmark{1}, Michel Fich\altaffilmark{1}, Peter Ade\altaffilmark{2}, Dan Bintley\altaffilmark{3}, Ed Chapin\altaffilmark{4, 5, 3}, Antonio Chrysostomou\altaffilmark{3}, James S.\,Dunlop\altaffilmark{6},  Andy Gibb\altaffilmark{5}, Jane S.\,Greaves\altaffilmark{7}, Mark Halpern\altaffilmark{5}, Wayne S.\,Holland\altaffilmark{8,6}, Rob Ivison\altaffilmark{8,6}, Tim Jenness\altaffilmark{9,3}, Ian Robson\altaffilmark{8}, Douglas Scott\altaffilmark{5}}

\altaffiltext{1}{Department of Physics \& Astronomy, University of Waterloo, Waterloo, Ontario, Canada, N2L 3G1}
\altaffiltext{2}{Cardiff School of Physics and Astronomy, Cardiff University, Cardiff, United Kingdom}
\altaffiltext{3}{Joint Astronomy Centre, 660, N. A'ohoku Place, University Park, Hilo, HI 96720, United States}
\altaffiltext{4}{XMM SOC, ESAC, Apartado 78, 28691 Villanueva de la Canada, Madrid, Spain}
\altaffiltext{5}{Department of Physics \& Astronomy, University of British Columbia, 6224 Agricultural Road, Vancouver BC V6T 1Z1, Canada}
\altaffiltext{6}{Institute for Astronomy, University of Edinburgh, Royal Observatory, Blackford Hill, Edinburgh EH9 3HJ}
\altaffiltext{7}{School of Physics and Astronomy, University of St Andrews, North Haugh, St Andrews, Fife KY16 9SS, UK}
\altaffiltext{8}{UK Astronomy Technology Centre, Royal Observatory, Blackford Hill, Edinburgh EH9 3HJ}
\altaffiltext{9}{Department of Astronomy, Cornell University, Ithaca, NY 14853}

\begin{abstract}
We mapped five massive star forming regions with the SCUBA-2 camera on the James Clerk Maxwell Telescope (JCMT). Temperature and column density maps are obtained from the SCUBA-2 450 and 850 $\mu$m images. Most of the dense clumps we find have central temperatures below 20 K with some as cold as 8 K, suggesting that they have no internal heating due to the presence of embedded protostars. This is surprising, because at the high densities inferred from these images and at these low temperatures such clumps should be unstable, collapsing to form stars and generating internal heating. The column densities at the clump centres exceed 10$^{23}$ cm$^{-2}$, and the derived peak visual extinction values are from 25-500 mag for $\beta$ = 1.5-2.5, indicating highly opaque centres. The observed cloud gas masses range from  $\sim$ 10 to 10$^{3}$ M$_{\odot}$. The outer regions of the clumps follow an $r^{-2.36\pm0.35}$ density distribution and this power-law structure is observed outside of typically 10$^{4}$ AU. All these findings suggest that these clumps are high-mass starless clumps and most likely contain high-mass starless cores. 
\end{abstract}

\keywords{\HII\ regions, dust, extinction, ISM: clouds, stars: formation, submillimeter: ISM}

\section{INTRODUCTION}
The dust grains in clouds associated with massive star-forming regions absorb the short wavelength radiation from these stars, heat up, and then re-radiate at far-infrared (FIR) and sub- millimeter wavelengths. For this reason, clouds near \HII\ regions are good sources of thermal emission. Since thermal radiation from dust is optically thin at FIR and sub-millimetre wavelengths, it is a good tracer of the temperature, density and gas mass of the clouds and these properties are significant in characterizing the early stages of star formation. Gas masses are usually determined from CO or dust continuum observations. However CO is optically thick, its column density subject to an \textit{X}-factor \citep{Dickman78} that is not known (the Milky Way value is usually assumed) and is probably sensitive to the local metallicity or background UV field \citep{Dishoek88, Leory09, Bolatto13}. 
Dust mass are usually derived from (sub)millimetre observations on the R-J side of the SED, which are sensitive to dust temperatures down to 5-10 K, and in addition are probably also dependent on the metallicity.

Values of dust temperatures and masses estimated from dust emission are subject to the adopted values of the dust emissivity index, $\beta$. Values of $\beta$ in molecular clouds reflect different physical and chemical properties of the clouds, such as the grain size, structure, composition and environment. At far-infrared wavelengths, the preferred value of $\beta$ in the diffuse interstellar medium is $\sim$ 2 \citep{draine_lee84} and this is the commonly assumed value of $\beta$ at long wavelenghts. However, a number of observations of cold cores \citep{Shirley05, PlanckC11, Shirley11} detected sub-millimetre dust emissivity indices greater than 2. In cold, dense regions of molecular clouds, $\beta$ can be as steep as $\sim$ 3.7 \citep{Kuan96} due to the accretion of ice mantles and grain growth. Towards circumstellar disks, the value of $\beta$ can be as flat as $\sim$1 \citep{Beck91}. Values of dust emissivity indices as low as $\sim$ 1 are also observed towards Galactic cores with internal heating sources, perhaps caused by sightline variations in temperature \citep{Juvela11}. Observations of some Class 0 young stellar objects \citep{Kwon09} and studies of circumstellar disks around Class II protostars, T Tauri stars \citep{Ricci10b, Ricci10a, Ubach12}, young brown dwarfs \citep{Ricci12} in low mass star-forming regions suggest even lower values of $\beta$, less than unity.

Although an accurate measurement of $\beta$ is critical in determining the physical properties of the clouds, the relationship between $\beta$ and dust temperature ($T_{\rm{d}}$) is still under debate. Several studies suggest that $\beta$ and $T_{\rm{d}}$ are anti-correlated \citep{Dupac03, Desert08, Paradis10}. However, such an inverse relation between $\beta$ and $T_{\rm{d}}$ resulting from least-square fits to flux densities is sensitive to noise and/or temperature variations along the line-of-sight \citep{Shetty09a, Shetty09b}. A novel hierarchical Bayesian technique developed by \citet{Kelly12} finds a weak positive correlation between $\beta$ and $T_{\rm{d}}$, but they use an unphysical assumption of isothermal grey body SED parameters. In their comprehensive study to examine the validity of five different potential methods on the shape of the $\beta$-$T_{\rm{d}}$ relation, \citet{Juvela13} found an inverse $T_{\rm{d}}$-$\beta$ correlation with all the methods, and they noticed a lower bias especially when using the Bayesian method.

Numerous studies have shown that a substantial, but not very well constrained, fraction of the mass of the molecular material is in very cold clouds---at least as measured by the dust---and is not observable except at sub-millimetre wavelengths. One of the best studied GMC \HII\ regions is the Orion Molecular Cloud (OMC)---a moderately luminous region ($\sim$5$\times$10$^{5}$ L$_{\odot}$ in the FIR), which is located at a distance of 420 pc. Using Submillimeter High Angular Resolution Camera (SHARC), \citet{Lis98} observed Orion A molecular cloud and their studies revealed an average dust temperatures of 17$\pm$4 K. Dust at temperatures down to 15 K is identified in a clump associated with Orion A \citep{Mookerjea00} from far-infrared observations at 138 and 205 $\mu$m wavelengths. Mookerjea et al. also found signatures of emission at temperatures between 15 and 20 K, corresponding to cold dust, in a number of Orion clumps and virtually all of the dust mass is in this cold component. Orion is extensively analyzed using SCUBA by Johnston et.al in a series of papers: the internal dust temperature of the clumps in Orion B molecular cloud is in the range 20-40 K \citep{Johnstone01}, whereas the best-fit temperature in Orion B South molecular cloud is 18$\pm$4 K \citep{Johnstone06a}, and in Orion A clumps, this is $\sim$22$\pm$5 \citep{Johnstone06b}. However, there is no overall census of the relative amounts of cold dust and warmer dust in clouds such as the OMC. One of the reasons for this is that it has been difficult to observe the extended emission, which is very extended in the case of the OMC.

Cold dust is observed in both low- and high-mass star-forming regions and the lowest temperatures found toward these regions differ only by a couple of Kelvins. The dust temperatures measured toward low-mass star-forming regions are found to be as low as $\sim$ 7 K in a number of clouds \citep{Evans01}, below $\sim$ 10-11 K in the pre-stellar cores of  $\varrho$ Oph main cloud \citep{Stam07}, and $\sim$ 7 K in TMC-1C \citep{Schnee10}. Over the last couple of years, a number of papers have focused on the physical properties of dust near low and high-mass star forming regions in the 250--500 $\mu$m wavelength regime and have found dust temperatures as low as 10 K in molecular clouds. From \textit{Herschel Space Observatory} \citep{Pilbratt10} data of a collection of low-mass cold, dense globules, \citet{Laun13} obtained temperatures in the range 8-12 K at the interiors. The Balloon-borne Large Aperture Submillimeter Telescope (BLAST; \citet{Pascale08}) observations of an OB forming complex, Cygnus-X, show some sources with temperatures as cold as $\sim$ 10 K \citep{Roy11a}. ``Herschel imaging survey of OB Young Stellar objects (HOBYS)" program \citep{Motte10} have provided unprecedented details of cold dust associated with \HII\ regions. Using data from \textit{Herschel} and other short and long wavelength data, the dust temperature map of the Galactic bubble \HII\ region RCW 120, derived by \citet{Anderson10} shows temperatures down to $\sim$10 K towards local infrared dark clouds (IRDCs). Vela-C, which is a site for both low- and high-mass star formation, shows warmer protostellar sources, with mean dust temperatures of 12.8 K, compared to starless sources with mean dust temperatures of 10.3 K \citep{Giannini12}. However, not all of these estimates of temperatures can be directly compared because of differences in wavelength coverage, modelling assumptions, and statistical treatments. In particular, the detection of very cold dust requires observations at longer wavelengths. Dust at temperatures below 10 K will have a peak in emission at wavelengths longer than 300 $\mu$m.

Observations of star formation in other galaxies is dominated by regions where clusters of massive stars heat their surrounding region. The properties measured for these regions, such as temperature, luminosity, mass, and star formation efficiency are determined by the interactions of the massive stars and the molecular clouds near them. Understanding how to turn the observations into measured properties is highly dependent on what we have learned about star formation in similar massive star-forming regions in the Milky Way. However only a handful of such regions have been well studied and all of these studies suffer from a number of difficulties. To probe such regions, what we really require is higher spatial resolution and longer wavelength observations. In this paper, we use the new SCUBA-2 camera \citep{Holland13}, at 450 $\mu$m and 850 $\mu$m wavelengths, to observe a carefully selected sample of massive star-forming regions in the outer Galaxy in order to overcome these problems.

In this paper, we calculate physical properties such as the temperature, density, and mass of the clouds associated with five massive star forming regions in the outer Galaxy. The clouds are found within 15$\arcmin$ of \HII\ regions, generally with no other significant concentration of interstellar material within many tens of arcminutes. We show the existence of very cold dust, with grain temperatures as low as 7.7 K near to very hot massive star-forming regions. 
 
\section{OBSERVATIONS AND DATA REDUCTION}
We used the SCUBA-2 instrument to observe five regions of massive star formation, centered on the \HII\ regions: (1) S148; (2) S156; (3) S159; (4) S305; and (5) S254. Some of these regions only contain one \HII\ region; some contain as many as five \HII\ regions. These \HII\ regions are located in the outer Galaxy and  there is very little confusion and contamination with foreground or background cold cores along the line of sight. The distances to our target objects are all well measured and they are a few kiloparsecs (2.5 to 5.6) away from the Sun. The target objects are all very bright and could be easily mapped using SCUBA-2, since their angular sizes and separations between their component clumps are small. All of these objects have many observations at other wavelengths from optical to centimetre. The physical properties of all the \HII\ regions are summarized in Table \ref{HII}. Uncertainties on the  linear diameter range from $\sim$ 5 to 27\% with a mean uncertainty of $\sim$ 13\%, and are primarily due to the uncertainty in the distance to the \HII\ region.  

\begin{table}[ht]
\caption{Properties of the chosen \HII\ regions: \HII\ region name; distance; angular diameter; physical diameter; and exciting star.}
\label{HII}
\begin{center}
\leavevmode
\begin{tabular}{l c c c c c c} \hline \hline                
Name & RA  & Dec &$d$  &Diameter &Diameter & Exciting star\\ 
	    & (J2000)&(J2000)& (kpc) &($\arcsec$) & (pc) & \\
\hline
\\
S148 complex\\
S147		&  	22$^{\rm{h}}$55$^{\rm{m}}$40.6$^{\rm{s}}$ & 58$\degree$28$\arcmin$02.0$\arcsec$ &5.50$\pm$0.89	&90		&2.44	&	A4 V\tablenotemark{2} ?\\
S148    	&  	22$^{\rm{h}}$56$^{\rm{m}}$05.7$^{\rm{s}}$ & 58$\degree$30$\arcmin$55.4$\arcsec$ &5.60$\pm$0.3	&90 		&2.44	&	O8 V\tablenotemark{2} \\ 
S149		&  	22$^{\rm{h}}$56$^{\rm{m}}$22.3$^{\rm{s}}$ & 58$\degree$31$\arcmin$57.1$\arcsec$ &5.60$\pm$0.3	&120		&3.30	&	BO V\tablenotemark{2} \\
\\
S156 complex\\
BFS15  	&  	23$^{\rm{h}}$04$^{\rm{m}}$42.6$^{\rm{s}}$ & 60$\degree$04$\arcmin$56.1$\arcsec$ &3.0$\pm$0.7	 	& 61		&0.88	&	\nodata	 	\\
S156    	&  	23$^{\rm{h}}$05$^{\rm{m}}$10.6$^{\rm{s}}$ & 60$\degree$14$\arcmin$57.6$\arcsec$ & 2.87$\pm$0.75 	& 40  	&0.56	&	O8 V\tablenotemark{1} \\
BFS18	&  	23$^{\rm{h}}$05$^{\rm{m}}$47.3$^{\rm{s}}$ & 60$\degree$24$\arcmin$06.8$\arcsec$ & 3.0$\pm$0.7	& 59		&0.86	&	\nodata	 	\\
\\
S159    	&  	23$^{\rm{h}}$15$^{\rm{m}}$31.2$^{\rm{s}}$ & 61$\degree$06$\arcmin$55.5$\arcsec$ & 2.97$\pm$0.31    &15 		&0.22 	&	O9 V\tablenotemark{1} \\
\\
S254 complex\\
S254 	&  	06$^{\rm{h}}$12$^{\rm{m}}$20.3$^{\rm{s}}$ &18$\degree$02$\arcmin$32.5$\arcsec$ &2.46$\pm$0.16	&540		&6.44	&	O9.6 V\tablenotemark{4}	\\
S255		&	06$^{\rm{h}}$13$^{\rm{m}}$09.5$^{\rm{s}}$ &17$\degree$58$\arcmin$40.6$\arcsec$ &2.46$\pm$0.16	&300		&3.58	&	B0.0 V\tablenotemark{4}	\\
S256		&	06$^{\rm{h}}$12$^{\rm{m}}$39.3$^{\rm{s}}$ &17$\degree$56$\arcmin$48.1$\arcsec$ &2.46$\pm$0.16	&120		&1.43	&	B0.9 V\tablenotemark{4}	\\
S257		&	06$^{\rm{h}}$12$^{\rm{m}}$49.1$^{\rm{s}}$ &17$\degree$58$\arcmin$46.1$\arcsec$ &2.46$\pm$0.16	&240		&2.86	&	B0 V\tablenotemark{5,6}\\
S258		&	06$^{\rm{h}}$13$^{\rm{m}}$33.6$^{\rm{s}}$ &17$\degree$55$\arcmin$39.7$\arcsec$ &2.46$\pm$0.16	&59		&0.7		&	B1.5 V\tablenotemark{4}	\\
\\
S305  	&  	07$^{\rm{h}}$30$^{\rm{m}}$07.0$^{\rm{s}}$ & -18$\degree$31$\arcmin$33.5$\arcsec$&5.20$\pm$1.40	&420 	&10.6	&	O8.5 V, O9.5 V\tablenotemark{3}\\
\\
\hline
\end{tabular}
\tablenotetext{1}{\citep{Rus07}}
\tablenotetext{2}{\citep{Crampton78}}
\tablenotetext{3}{\citep{Russeil95}}
\tablenotetext{4}{\citep{Fazio08}}
\tablenotetext{5}{\citep{Pismis76}}
\tablenotetext{6}{\citep{Moffat79}}
\end{center}
\end{table}

\subsection{SCUBA-2 data}
We observed the area around S148, S156, S159, S305, and S254 as part of the SCUBA-2 guaranteed time project M11BGT01, primarily in weather band 1 and 2, at various times between November 14, 2011 and March 14, 2012. The mean frequency of the SCUBA-2 filters are 665 GHz at 450 $\mu$m waveband and 349.5 GHz at 850 $\mu$m waveband \href{http://www.jach.hawaii.edu/JCMT/continuum/scuba2/filter/}{(see SCUBA-2 filter specifications)} with the half power beam widths of the band pass filters, corresponding to $\lambda/d\lambda$ $\sim$ 14 and 10, are 32 and 85 $\mu$m respectively, at 450 and 850 $\mu$m \citep{Holland13}. We used the rotating PONG \citep{Kackley10} method to map our target objects: the first three objects were observed in the PONG900 mode, which produced an image approximately 15$\arcmin$ $\times$ 15$\arcmin$ in size, and in order to capture the extended features, the last two observations made use of the 30$\arcmin$ $\times$ 30$\arcmin$ PONG1800 mapping mode. For some of the observations the beam size was observed to be larger than the nominal FWHM of 13.0 $\arcsec$ (850 $\mu$m)  and 7.9 $\arcsec$ (450 $\mu$m) \citep{Dempsey13}: the JCMT beam sizes were 14.2$\arcsec$ (850 $\mu$m) and 9.4$\arcsec$ (450 $\mu$m) for the first three observations, whereas the beam sizes were 14.0$\arcsec$ (850 $\mu$m) and 7.5$\arcsec$ (450 $\mu$m) for the last two observations.

The SCUBA-2 data were reduced using the \textit{Starlink} SMURF (version 1.3.8---for the first three observations and 1.3.10---for the last two observations) and the iterative map maker \citep{Chapin13}, which is part of the sub-millimetre user reduction facility SMURF \citep{JennessT2011} software package. We followed the recommended ``best practice'' \href{http://www.starlink.ac.uk/docs/sc21.htx/sc21.html}{(Thomas, H. S., 2013, Starlink Cookbook 21, Joint Astronomy Centre)} and used an externally generated zero mask to constrain the map-maker and prevent negative bowling.

The final maps were calibrated using flux calibration factors (FCFs) obtained from planet data for the respective observation dates. We derived the FCF using SMURF and the pipeline for combining and analyzing reduced data, PICARD \citep{JennessT08}\footnote{http://www.oracdr.org/oracdr/PICARD}, which is part of the ORAC-DR \citep{oracdr} pipeline and has the same infrastructure. The FCFs for S156 were derived at both wavelengths using the primary calibrator Mars on the night of the science observation; Uranus was also available as a primary calibrator. Due to bad focusing and atmospheric effects in the Uranus observation (seen primarily at 450- $\mu$m), which was taken before the period of the science observation in the early evening hours of the observation date, we selected a midnight Mars observation to derive the FCF. In addition, there were standard peak FCF values available from nearly 500 observations at both wavelengths \citep{Dempsey13}. The standard FCFs corresponding to the \textit{Starlink} SMURF version at the time of data reduction were (556$\pm$45) Jy beam $^{-1}$ pW $^{-1}$ and (606$\pm$55) Jy beam $^{-1}$ pW $^{-1}$ at 850 and 450 $\mu$m, respectively. The values of the FCFs used for calibrating S156 are 545 Jy beam $^{-1}$ pW $^{-1}$ at 850 $\mu$m and 563 Jy beam $^{-1}$ pW $^{-1}$ at 450 $\mu$m, and these are within the uncertainty limits of the standard values. The FCFs for S159 and S148, which were observed on the same day, obtained from both Mars and Uranus, displayed large variations from the standard value, especially at 450 $\mu$m. We therefore adopted the standard FCFs to calibrate S159 and S148 data.

In finding the FCFs of S305 and S254, we considered two different observations of the planet Mars acquired on the same night. We adopted the FCF derived from the early night Mars observation for calibrating the short wavelength data. For calibrating the long wavelength data, we chose a Mars observation near midnight. The values of the FCFs used to calibrate S305 and S254 data at 850 and 450 $\mu$m wavelengths are, respectively, 566 and 565 Jy beam $^{-1}$ pW $^{-1}$. The effect of FCF uncertainty on the final fluxes is less than 5\% at 850 $\mu$m and 10\% at 450 $\mu$m \citep{Dempsey13}.

\subsection{Role of Background on SCUBA-2 Data}
We chose data from one of the objects observed, S305 (the object with the most widespread emission structures), to examine in detail the effects of background emission on the measurements made from the images. We measured the flux density in same-sized apertures at numerous positions showing no discreet emission that might be due to real objects. When we increase the distance from the centre of the image, the mean background flux densities approach positive values. At the edges of the maps, especially near 25$\arcmin$, the background emission is the highest and is most likely due to the reduced integration time. The error retains a constant value up to around 20$\arcmin$ from the map centre and then begins to increase out to the edge. The mean background values at 450 $\mu$m show much more variability than the corresponding values at 850 $\mu$m. The background standard deviation, however, maintains a constant value up to around 20$\arcmin$ from the map centre, and then increases up to the map boundary. The mean background value of the S305 field ($\sim$--0.003 Jy beam $^{-1}$ at 850 $\mu$m \& $\sim$--0.009 Jy beam $^{-1}$ at 450 $\mu$m) is not significantly different from zero, relative to the noise ($\sim$0.02 Jy beam $^{-1}$ at 850 $\mu$m \& $\sim$0.06 Jy beam $^{-1}$ at 450 $\mu$m). Thus, on average, the background emission values are close to zero and therefore, we did not take into account the contribution due to background in the S305 source emission.

In all the target fields of our study, the contribution from the mean value of the background on the source flux is computed to be very low and is neglected in further calculations. Typical on-source signal, especially in the fainter extended regions of S305, is (0.12$\pm$0.04) Jy beam $^{-1}$ at 850 $\mu$m and (0.36$\pm$0.10) Jy beam $^{-1}$ at 450 $\mu$m. Note that the absolute value of the mean background emission in S305 is a factor of around 40 (at 850 and 450 $\mu$m) times smaller than the faint extended on-source emission. For all the target objects, the ratio of the faintest on-source signal to the mean background, on average, is around 41 at 850 $\mu$m and 17 at 450 $\mu$m. The total flux due to this offset in the background may be significant for a measure of the integrated flux from these objects, but our main result, the low temperatures of the centres of the clouds, depends on the ratios of surface brightness.

\section{PROPERTIES OF THE DUST DERIVED FROM THE IMAGES}
Since dust emission at sub-millimetre wavelengths is optically thin, it is a good tracer of the masses of the clouds. In order to estimate temperature, density, and mass, we employed the surface brightness maps of the clouds obtained from SCUBA-2 at 450 $\mu$m and 850 $\mu$m wavelengths. In this section we describe the clouds' physical properties derived using the observed sub-millimetre dust emission maps. The properties are dependent on environmental and instrumental factors, such as calibration, line contamination, knowledge of $\beta$, effect of foreground/background and temperature variations along the line of sight, all of which are examined at the end of Section \ref{anal}. and discussed in Section \ref{dis}.

\subsection{Dust Emission Morphology}
The five observed SCUBA-2 maps at  450 and 850 $\mu$m are displayed in Figures \ref{S148_panel}(a) to \ref{S305_panel}(a) and \ref{S156_panel}(b) to \ref{S305_panel}(b). Hereafter, north is up and east is to the left, with map coordinates in J2000. Due to large-scale sky variations in the 450 $\mu$m map of S148, we discarded this map from further analysis. A long bar like cloud whose strongest emission is located at 22h 56m 45.8s RA and +58$\degree$ 30$\arcmin$ Dec, aligned in the north-south direction, is visible in the south-east direction of the cloud associated with the \HII\ region S148. A third cloud is present in the S148 field at a distance of $\sim$6$\arcmin$ 45$\arcsec$ from the map centre to the north. Please note that there are no visible optical counterparts for these two unknown clouds in the DSS red image, and therefore these two clouds are most likely some dark clouds along the line of sight or associated with the centre cloud. Since the identity and the distances to these clouds are uncertain, we are not taking these two clouds into account in further studies. In the \HII\ region clouds under discussion, it is apparent that the 450 $\mu$m maps resemble their 850 $\mu$m counterparts. The maps presented here are in units of Jy beam$^{-1}$ and to convert these values to units of MJy sr$^{-1}$ (the other commonly used units), one has to multiply by 185 at 850 $\mu$m and 426 at 450 $\mu$m for S159, S148, and S156. For S305 and S254, the Jy beam$^{-1}$ values need to be multiplied by 192 at 850 $\mu$m and 668 at 450 $\mu$m to obtain corresponding values in MJy sr$^{-1}$.

The observed characteristics of the clouds in the vicinity of the individual \HII\ regions are thoroughly discussed in \citet{Gopika12} and are tabulated in Table \ref{obs}, where the peak positions and flux densities for each object are given. The uncertainties in the peak flux densities are primarily due to uncertainties in the flux calibration factor: i.e., $\sim$ 5\% at 850 $\mu$m and 10\% at 450 $\mu$m \citep{Dempsey13}. Note that many of these images show more than one area of significant emission. Some of these are more closely associated with other nearby \HII\ regions and we have indicated these with the appropriate names in this Table. Some of these emission features are roughly circular and isolated and later in this paper we identify them as ``clumps'' and analyse their properties. We presume that these clumps are gravitationally bound collections of cores and interferometric observations may reveal their fragmentation into individual cores at arc sec resolution. Throughout the paper ``cloud'' is a general term, that in addition to clumps, refers to less symmetric objects (although some of those might be more correctly described as filaments).  


\begin{table}
\caption{Observed parameters: clouds associated with \HII\ regions; peak positions; 850 and 450 $\mu$m peak flux densities of the clouds detected in the SCUBA-2 observations of Galactic \HII\ region complexes; and dust temperatures, ${{T_{\rm{c}}}_{\rm{d}}}$ (for $\beta=2$), which comes from the peak 450 and 850 $\mu$m emission. See Figures \ref{S148_panel} to \ref{S305_panel} for maps of these fields.}
\label{obs}
\begin{center}
\leavevmode
\begin{tabular}{l c c c c c} \hline \hline                
Source & RA  & Dec & $S^{\rm{peak}}_{850}$  & $S^{\rm{peak}}_{450}$ & ${{T_{\rm{c}}}_{\rm{d}}}$ \\ \\ 
	    & (J2000)&(J2000)& (Jy beam$^{-1}$) & (Jy beam$^{-1}$)&(K) \\
\hline
S148    	&  	22$^{\rm{h}}$56$^{\rm{m}}$17.2$^{\rm{s}}$ & 58$\degree$31$\arcmin$01.0$\arcsec$  & 0.50 & \nodata	& \nodata	\\ 
S156    	&  	23$^{\rm{h}}$05$^{\rm{m}}$08.9$^{\rm{s}}$ & 60$\degree$14$\arcmin$48.9$\arcsec$  & 1.09 & 3.49	&	11.5	\\
BFS15  	&  	23$^{\rm{h}}$05$^{\rm{m}}$24.6$^{\rm{s}}$ & 60$\degree$08$\arcmin$12.8$\arcsec$  & 2.09 & 5.48	&	9.5	\\
S156NE  	&  	23$^{\rm{h}}$06$^{\rm{m}}$18.7$^{\rm{s}}$ & 60$\degree$16$\arcmin$17.8$\arcsec$  & 0.35 &  \nodata	& \nodata	 \\ 
S159    	&  	23$^{\rm{h}}$15$^{\rm{m}}$30.7$^{\rm{s}}$ & 61$\degree$07$\arcmin$14.8$\arcsec$  & 2.34 & 9.03	& 	14.3		\\
S254N 	&  	06$^{\rm{h}}$12$^{\rm{m}}$53.6$^{\rm{s}}$ & 18$\degree$00$\arcmin$27.7$\arcsec$   & 4.97 & 21.7 	&	24.6	\\
S254S1 	&  	06$^{\rm{h}}$12$^{\rm{m}}$53.8$^{\rm{s}}$ & 17$\degree$59$\arcmin$24.7$\arcsec$   & 5.33 & 23.4	&	23.4\\
S254S2	&	06$^{\rm{h}}$12$^{\rm{m}}$56.8$^{\rm{s}}$ & 17$\degree$58$\arcmin$34.0$\arcsec$   & 1.29 & 4.94	&	28.6\\
S305N 	&  	07$^{\rm{h}}$29$^{\rm{m}}$56.2$^{\rm{s}}$ & $-$18$\degree$27$\arcmin$54.2$\arcsec$  & 0.71 & 2.22	&	13.7\\
S305W1	&	07$^{\rm{h}}$30$^{\rm{m}}$00.3$^{\rm{s}}$ & $-$18$\degree$33$\arcmin$33.5$\arcsec$  & 0.33 & 0.84 	&	13.8\\
S305W2	&	07$^{\rm{h}}$30$^{\rm{m}}$00.4$^{\rm{s}}$ & $-$18$\degree$31$\arcmin$09.5$\arcsec$  & 0.26  & 0.67 	&	14.7\\
S305W3	&	07$^{\rm{h}}$30$^{\rm{m}}$00.6$^{\rm{s}}$ & $-$18$\degree$31$\arcmin$54.5$\arcsec$  & 0.37  & 1.01 	&	14.1\\
S305W4	&	07$^{\rm{h}}$30$^{\rm{m}}$00.9$^{\rm{s}}$ & $-$18$\degree$34$\arcmin$44.5$\arcsec$  & 0.18  & 0.53 	&	12.9\\
S305W5	&	07$^{\rm{h}}$30$^{\rm{m}}$04.8$^{\rm{s}}$ & $-$18$\degree$31$\arcmin$03.5$\arcsec$  & 0.29  & 0.84 	&	13.8\\
S305E1	&	07$^{\rm{h}}$30$^{\rm{m}}$11.8$^{\rm{s}}$ & $-$18$\degree$32$\arcmin$36.3$\arcsec$  & 0.38  & 1.04 	&	14.3\\
S305S  	& 	 07$^{\rm{h}}$30$^{\rm{m}}$16.5$^{\rm{s}}$ & $-$18$\degree$35$\arcmin$51.2$\arcsec$ & 1.10 & 3.82	&	16.7\\
S305E2  	& 	 07$^{\rm{h}}$30$^{\rm{m}}$16.7$^{\rm{s}}$ & $-$18$\degree$33$\arcmin$54.3$\arcsec$ & 0.13 & 0.37	&	14.5\\

\hline
\end{tabular}
\end{center}
\end{table}

\subsubsection{Temperature distribution}
One can derive a pixel-to-pixel dust colour temperature map of the clouds by taking the ratio of 450 to 850 $\mu$m observed surface brightness maps:
\begin{equation}
\frac{S_{450} ({Jy/450\hspace{1mm} beam})}{S_{850} ({Jy/850\hspace{1mm} beam})}= \left( \frac{850}{450}\right )^{3+\beta} \frac{\exp(17K/T_{\rm{d}})-1}{\exp(32K/T_{\rm{d}})-1}
\end{equation}
\citep{Kramer02}

Before constructing the ratio maps, we first smoothed the 450 $\mu$m SCUBA-2 map to the 850 $\mu$m resolution of the JCMT beam using a symmetrical Gaussian point spread function to roughly match the Gaussian main beams at both the wavelengths. The error beams, however, are not taken into account in matching the fluxes at two wavelengths. The flux ratio is modelled assuming that the measured flux density at both wavelengths is due to thermal emission from optically thin dust grains and is dependent on the dust emissivity index and temperature. In all the temperature maps, the distribution is derived by assuming a constant value for the dust emissivity index $\beta$. Note that the cloud S148 lacks a good detection at 450 $\mu$m, and so we were unable to provide a detailed dust temperature map for this cloud. To derive the dust physical properties, we considered a $\beta$ value of 2, which is often considered to be the standard value at sub-mm and mm wavelengths \citep{draine_lee84}. The effect of varying emissivity index is discussed in section \ref{anal} in order to investigate the effect of the assumed value of $\beta$ on the calculated values of the temperature, density, and mass. The variations in the dust temperature with the flux ratio values are dependent on the adopted values of $\beta$. If the 450/850 flux ratio is above $\approx$ 8.3, 11.5, 15.6 for $\beta =1.5, 2, 2.5 $, respectively, then no temperature can be found with the assumptions used here (e.g., uniform dust and dust opacity along the line of sight).

Dust colour (450/850) temperature maps of the clouds are calculated using a single dust emissivity index, $\beta=2.0$, and assuming that SCUBA-2 emission at 450 and 850 $\mu$m are solely originating from dust grains at a single effective temperature. The derived temperatures represent the average of the observed temperatures along the line of sight through the cloud depths [see Figures \ref{S156_panel}(c), \ref{S159_panel}(c), \ref{S254_panel}(c), and \ref{S305_panel}(c)] but with some additional weighting, of an unknown amount, on the hotter outer parts of each cloud.

There are just a few, almost random, points at the outer edges of each structure, where the S/N is low, that are well over 100 K and they may be just the result of noise. To improve our display of the interior regions, we have set the maximum in the colour tables for these figures at 80 K, except for S254, where we have set the maximum to 150 K since the temperatures are  much higher in the central regions of this complex. For most of the structures shown, this has resulted in two or three pixels ``disappearing'' at the noisy edges but in the case of S159, approximately 20 ``hot" pixels in the outer edge are not shown. In each cloud, a gradient in temperature is observed increasing from the cloud centre to the periphery.

Temperatures are not derived for the cloud associated with S148, due to large scale sky variations in the 450 $\mu$m map. We assume a uniform temperature of 10 K throughout S148 to derive the rest of the dust characteristics at sub-millimetre wavelengths. Most of the sources in the S156 field show dust temperature gradients [see Figure \ref{S156_panel}(c)]. The clump near the \HII\ region S156 is cold, with central temperatures $\sim$ 11.5 K, and most of the interior of this clump is below 20 K. The extended region of the clump S156, located in the south-east direction, is relatively warm---reaching on average $\sim$ 43 K with some pixels at the edge showing temperatures of $\sim$ 79 K. Note that the clump associated with the \HII\ region BFS15, located in the south, is showing a central temperature of 9.5 K and $\sim$ 28 K, on average, at the outer layers. The 450 $\mu$m emission of the cloud S156NE is weak ($\sim$ 1 Jy beam$^{-1}$) and uniform, while the 850 $\mu$m emission, although weak, does show a peak in emission. There are a couple of pixels near the centre of S156NE showing temperatures below 30 K, possibly due to noise in this generally low S/N region; most of the interior of S156NE shows temperatures of 30 to 60 K with an increase towards the outer edges. Temperature variations are high in the clump associated with S159, with some of the pixels at the main clump's outer boundary reaching around 80 K. However, the pixel at the centre with the lowest value of $T_{\rm{d}}$ is at 14.3 K. Outer parts of the clump retain a temperature between 30 and 60 K. All the fragments located in the south-west and north-east directions of the primary cloud show lowest temperatures of around 10 K. The minor fragments visible in the south-east direction are significantly warmer: the temperature of these regions are found to be above 20 K [see Figure \ref{S159_panel}(c)]. S159 is selected for more detailed analysis in section 3.2 and 3.3, as this is the largest, therefore best resolved, bright,  circularly symmetric object.

The centres of the clumps associated with the S254 complex, primarily S254N, S254S1, and S254S2, are warmer than the centres of clumps associated with other \HII\ regions in this study. The centres are at temperatures $\sim$ 23 K (S254S1), 25 K (S254N), 29 K (S254S2) and a couple of pixels at the exterior boundary show temperatures beyond 100 K [Figure \ref{S254_panel}(c)]. These warmer clumps are most likely associated with the presence of two hot \HII\ regions on either side of the clouds. This is the hottest environment in our sample, as these clumps are surrounded by not only these two much larger \HII\ regions, but there are also several other large \HII\ regions in the immediate vicinity. The cloud fragment in the south-east direction, at RA 06h 13m 29.2s and Dec +17$\degree$ 55$\arcmin$ 30.8$\arcsec$, connected to the \HII\ region S258, shows similar characteristics, with around 23 K dust at its centre. However, the cloud located in the south-east direction at RA 06h 13m 47.0s and Dec +17$\degree$ 54$\arcmin$ 54.5$\arcsec$ show a relatively colder central temperature of $\sim$ 11 K. The minimum values of dust temperatures in the clouds of S305 complex range from $\sim$ 12 K to 17 K. The clump at the south-east (S305S) is the warmest among all other nebulosities, with central temperatures of $\sim$ 17 K and grain temperatures reaching $\sim$ 46 K (on average) at the boundary. All other components of the complex are colder than S305S, with central temperatures ranging from $\sim$ 12 to 14 K [Figure \ref{S305_panel}(c)].

We identify 15 clouds with good S/N in both 450 and 850 $\mu$m wavelengths (listed in Table \ref{obs}). We find that 11 of these have central temperatures, ${{T_{\rm{c}}}_{\rm{d}}}$, below 15 K, and only 3 have ${{T_{\rm{c}}}_{\rm{d}}}$ above  20 K (typical uncertainties in ${{T_{\rm{c}}}_{\rm{d}}}$ are about 11\%). Temperatures below 20 K strongly suggests that there is no central heating source such as an embedded protostar. The average value of the dust temperatures at the centres of all the clouds associated with \HII\ regions is $\sim$ 16 K. The lowest dust temperature is at the centre of the clump BFS15 (9.5 K) and the warmest one is in the S254 complex, S254S2 (28.6 K). In all the objects the pixel with the lowest temperature can be driven by noise and to improve the estimate of the clump, we average over a 14$\arcsec$ diameter region and this typically produced a temperature $\sim$ 0.7 to 3.6 K higher. However, in the central clumps of S254 complex, this produced $\sim$ 7 to 25 K higher temperature.

Line-of-sight averaged dust temperature ($\langle T_{\rm{d}} \rangle_{los}$) profiles of all the clumps are derived by assuming that each clump has a spherical structure. The profile values are obtained by averaging the line-of-sight observed temperatures over concentric annuli of width 0.5 times the 850 $\mu$m FWHM centred on the clump, and with uncertainties in the temperature values representing the quadratic sum of the errors due to calibration and the standard deviation from the cylindrical symmetry at each radial annulus. The majority of the clumps have $\langle T_{\rm{d}} \rangle_{los}$ less than $\sim$ 28 K, with temperatures ranging from $\sim$ 13-28 K (in the centre) to $\sim$ 25-50 K (at the edges) except the clumps in the central regions of S254 complex, where the $\langle T_{\rm{d}} \rangle_{los}$ are higher, with $\sim$ 42-54 K at the centre and $\sim$ 90-100 K at the clump boundary.

The observed $\langle T_{\rm{d}} \rangle_{los}$ profile of one of the clump (near S159) for $\beta$=2 is shown in Figure \ref{S159_Tdr_cont_multibeta}(a) (red curve), where we examine the effect of line contamination and $\beta$ on our results (as discussed below). The observed $\langle T_{\rm{d}} \rangle_{los}$ distribution shows a rise in the temperature from the clump's centre to the outer envelope, indicating less input heating in the centre, for a constant value of the emissivity index. This feature in our target clumps is most likely attributed to the lack of any central heating source, since cores with increasing temperatures towards smaller radii represent cores that enclose central heating sources [e.g., \citet{Vtak00}]. However, in our clumps, there may be external radiation field, due to the exciting stars of the nearby \HII\ regions. The only clouds in our sample that would have higher central temperatures due to this are those with lower columns and/or higher external radiation fields, and both of these conditions appear to be true for the sources in the central filament of S254. The range of temperatures found in our study is slightly higher than that obtained from the low mass starless core TMC-1C in the Taurus molecular cloud: for a $\beta$ of 1.5, they observe dust temperatures ranging from 6 K to 15 K from the centre to the outer envelope \citep{Schnee05} whereas our study on S159 for a $\beta$ of 2 reveals somewhat higher dust temperatures, ranging from $\sim$ 14 K to 50 K from the centre to the outer envelope.

\subsubsection{Column Density Distribution}
We calculated the cloud optical depth assuming that the dust emission is optically thin at SCUBA-2 wavelengths. The 850 $\mu$m map (which is in units of Jy beam$^{-1}$) is selected to calculate the beam averaged optical depth. Using the 850 $\mu$m flux density, the beam area at 850 $\mu$m, the Planck function at 850 $\mu$m ($B_{850}$), and the 450/850 colour temperature found above, we can derive the optical depth at 850 $\mu$m ($\tau_{850}$):

\begin{equation}
\tau_{850} = \frac{F_{850}}{{\Omega_{\rm{b}}}B_{850}(T_{\rm{d}})}.
\end{equation}

Since thermal emission from dust at such long wavelengths is optically thin, $\tau_{850} << 1$, sub-millimetre dust emission can follow high column densities and the column density of hydrogen (H) from 850 $\mu$m observations can be derived directly from the optical depth using the formula

\begin{equation}
{{N_{\rm{H}}}_{850}}=\frac{\tau_{850}}{\kappa_{850} {\mu} {m_{\rm{H}}}}, 
\end{equation}

\noindent where $\kappa_{850}$ is the mass absorption coefficient, the opacity per unit mass column density, at 850 $\mu$m. We adopt $\kappa_{850}=0.01$ cm$^{2}$ g$^{-1}$ for a gas-to-dust mass ratio of 100 (opacity per unit dust mass, $\kappa_{850,\rm{d}}=1$ cm$^{2}$ g$^{-1}$) in the column density calculations of the target objects. The quantity $\mu$ (2.33) is the mean molecular weight of the cloud material, and $m_{\rm{H}}$ is the mass of a hydrogen atom. The measured column density is proportional to the flux at a particular wavelength, as well as the temperature of the dust. We also estimate the extinction $A_{\uppercase{V}}$ at each pixel position using the widely accepted Milky Way value of $R_{\uppercase{V}}$=3.1  in calculations; any change in $R_{\uppercase{V}}$, therefore, changes the value of visual extinction, $A_{\uppercase{V}}$. For example, if we adopt an $R_{\uppercase{V}}$ of 5.5, as found for denser star forming regions \citep{Mathis90, Draine03, Whittet2003}, in our calculations, then the visual extinction of the clouds will be modified by a factor $\sim$2.

\begin{figure}[ht!]
\begin{center}
\includegraphics[scale=0.65]{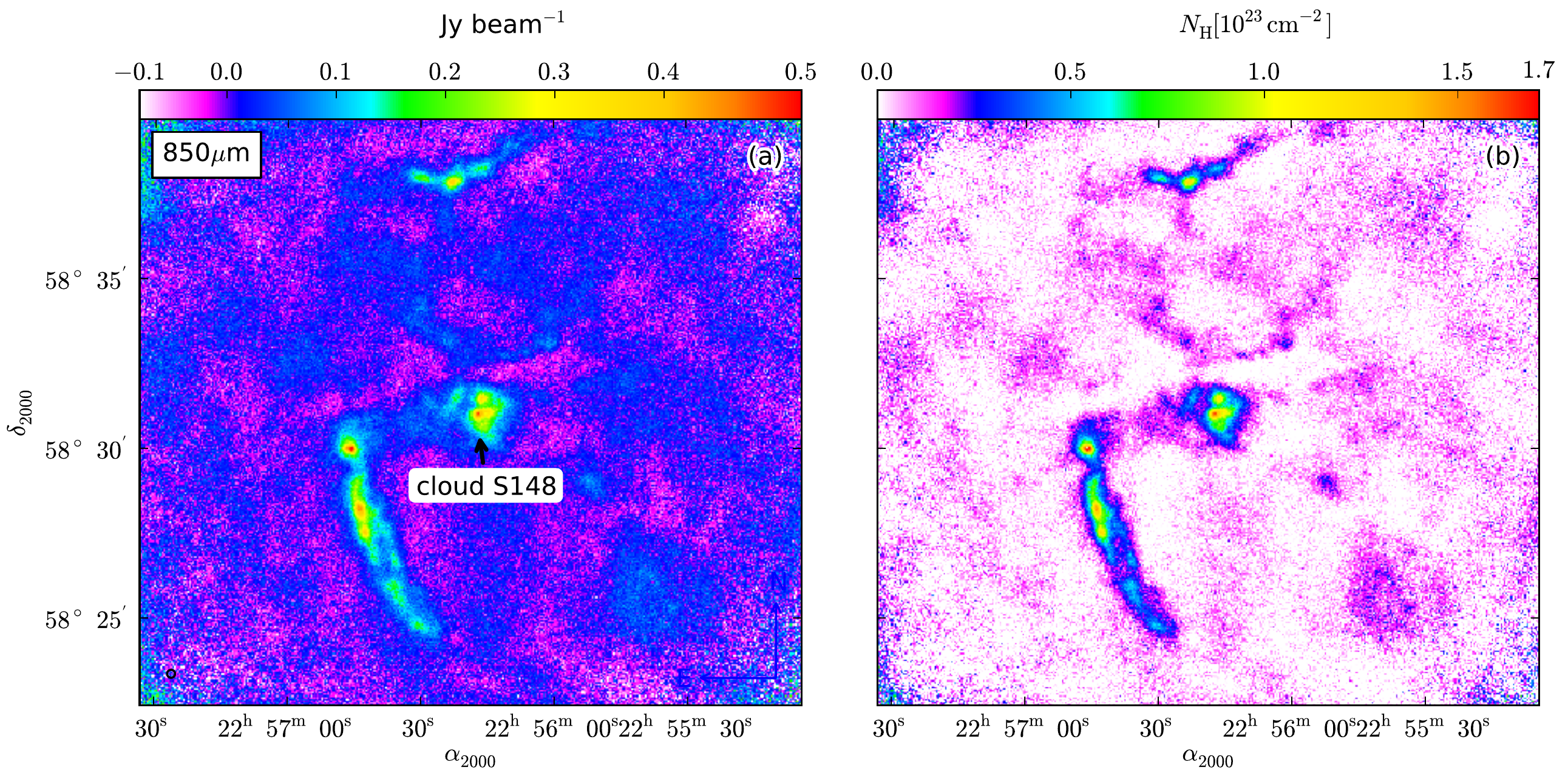} 
\end{center}
\caption{S148: (a) surface brightness map at 850 $\mu$m; and (b) column density map of S148 at 850 $\mu$m using temperature derived from assuming a $T_{\rm{d}}$ of 10 K.}
\label{S148_panel}
\end{figure}

\begin{figure}[ht!]
\begin{center}
\includegraphics[scale=0.60]{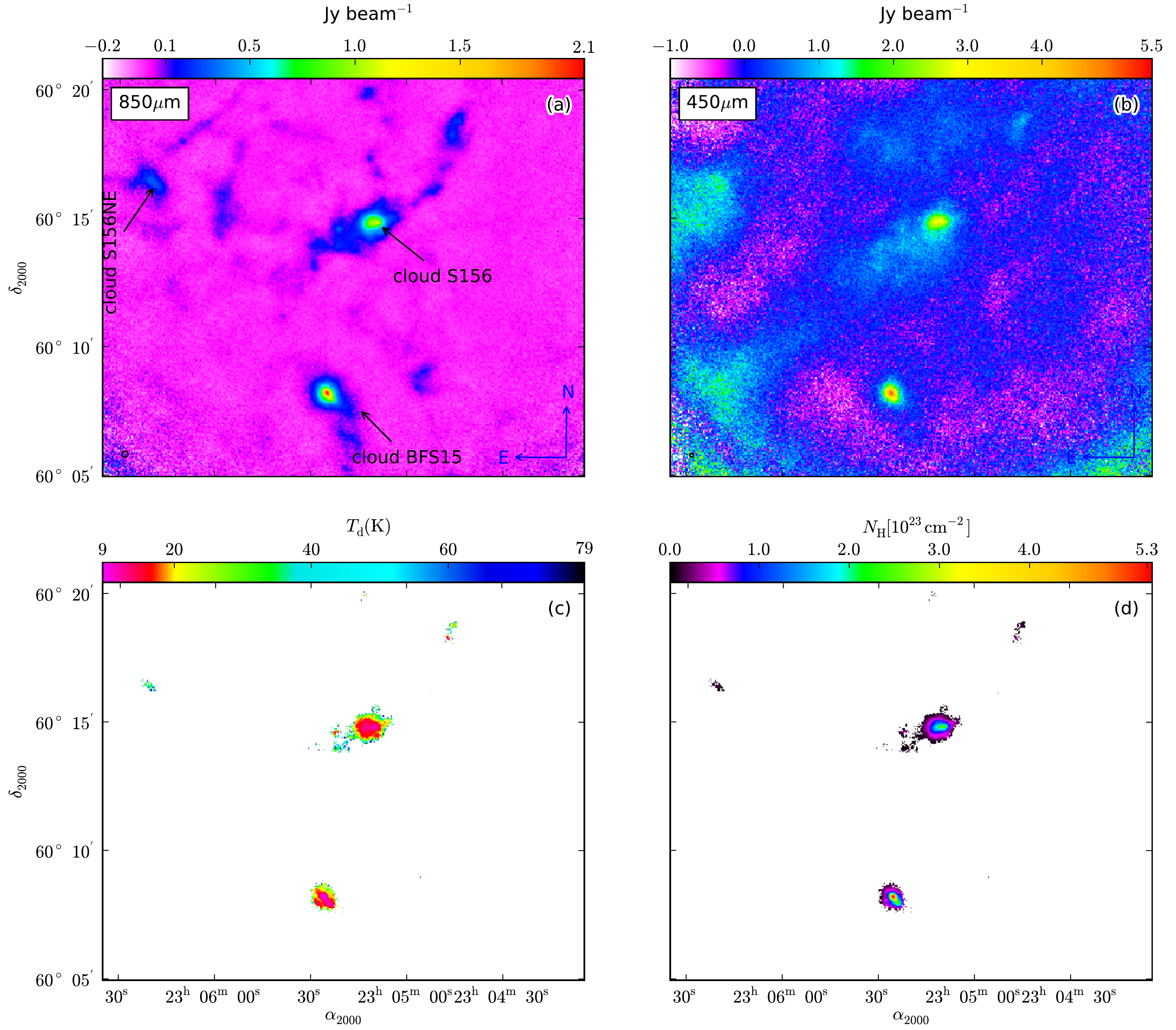}
\end{center}
\caption{S156: (a) surface brightness map at 850 $\mu$m; (b) surface brightness map at 450 $\mu$m; (c) dust temperature map; and (d) column density map.}
\label{S156_panel}
\end{figure}

\begin{figure}[ht!]
\begin{center}
\includegraphics[scale=0.54]{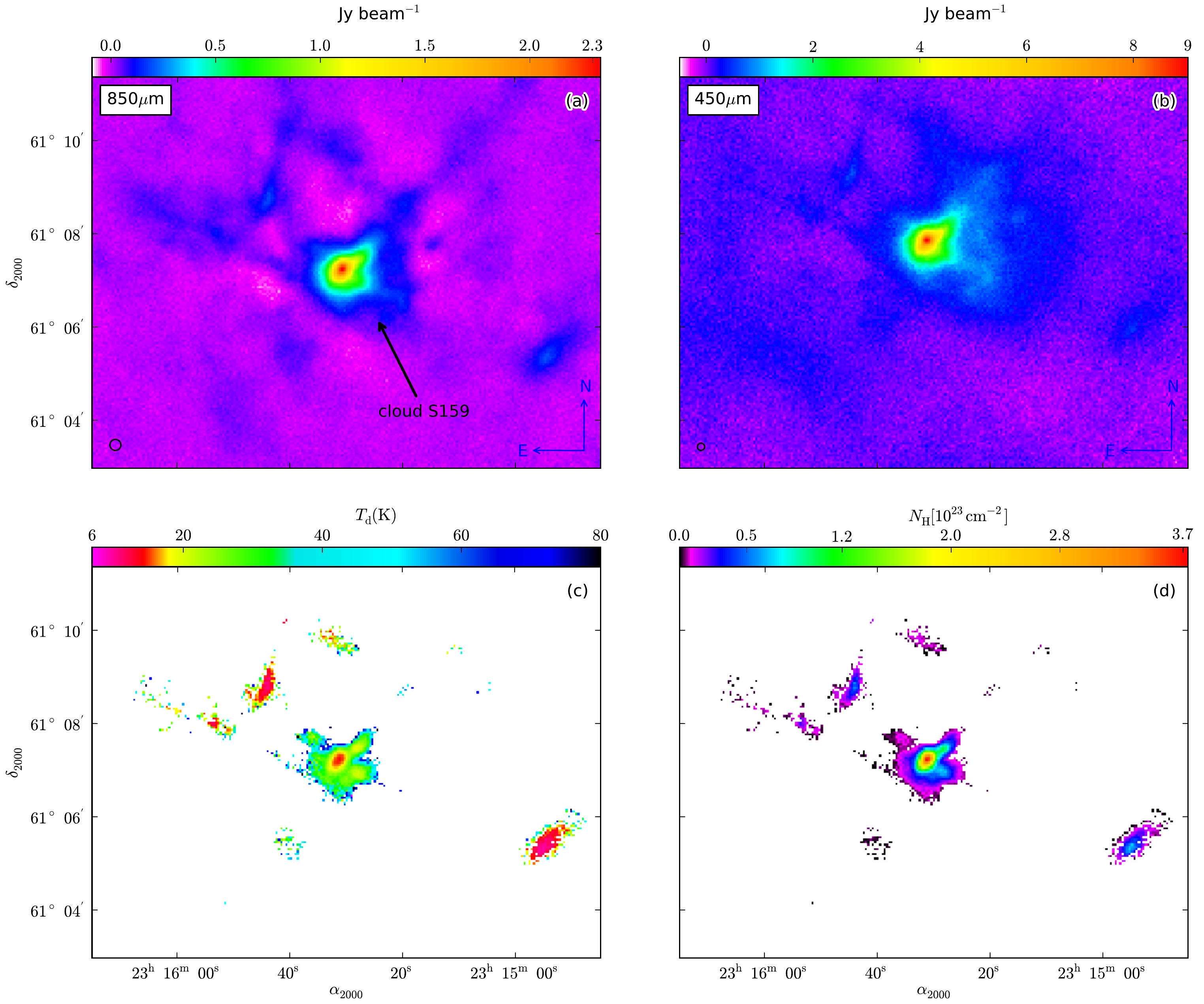}
\end{center}
\caption{S159: (a) surface brightness map at 850 $\mu$m; (b) surface brightness map at 450 $\mu$m; (c) dust temperature map; and (d) column density map.}
\label{S159_panel}
\end{figure}

\begin{figure}[ht!]
\begin{center}
\includegraphics[scale=0.55]{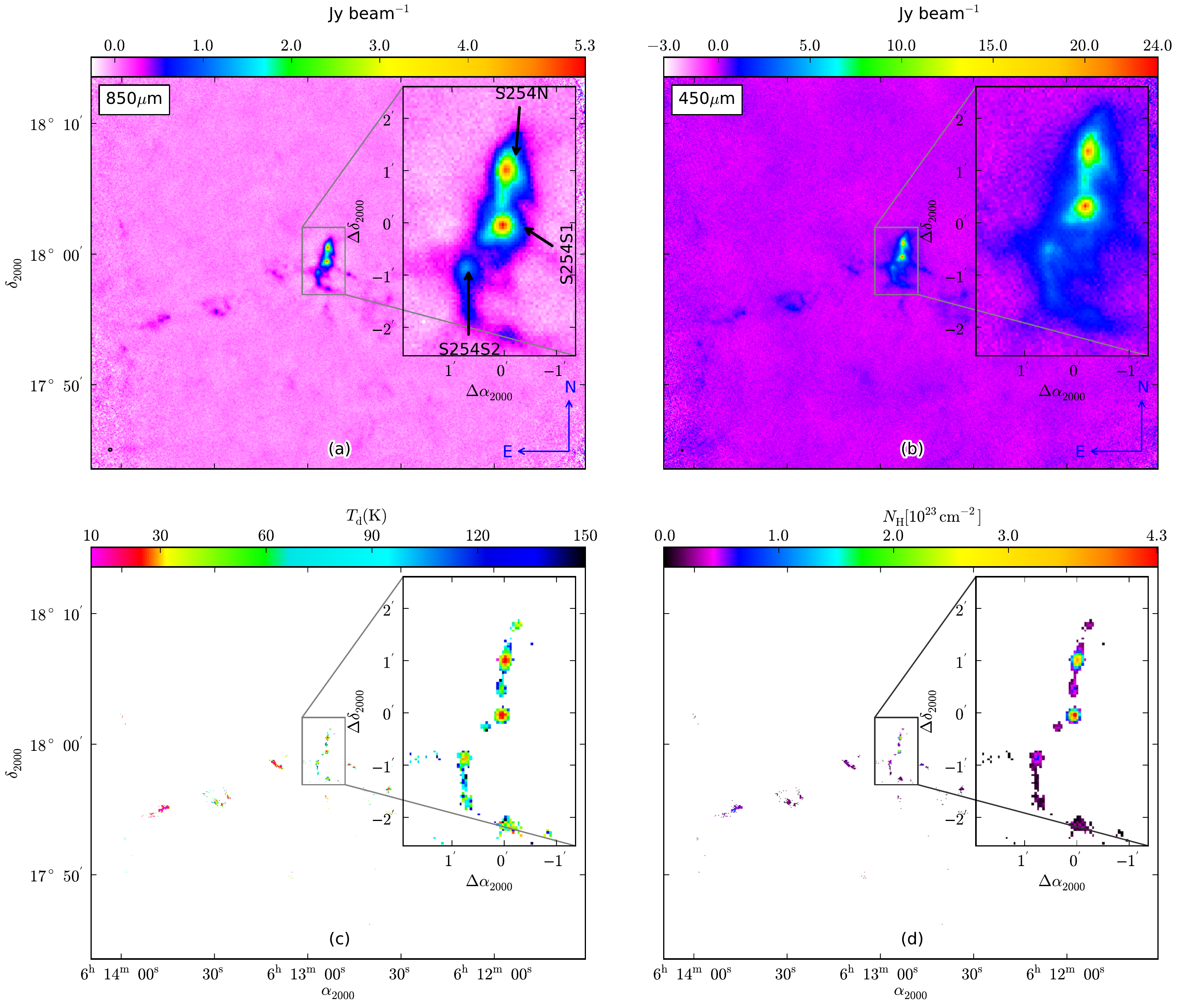}
\end{center}
\caption{S254: (a) surface brightness map at 850 $\mu$m; (b) surface brightness map at 450 $\mu$m; (c) dust temperature map; and (d) column density map. The centre for the inset map is at RA 06$^{\rm{h}}$12$^{\rm{m}}$53.9$^{\rm{s}}$ and Dec 17$\degree$59$\arcmin$24.7$\arcsec$ }
\label{S254_panel}
\end{figure}

\begin{figure}[ht!]
\begin{center}
\includegraphics[scale=0.54]{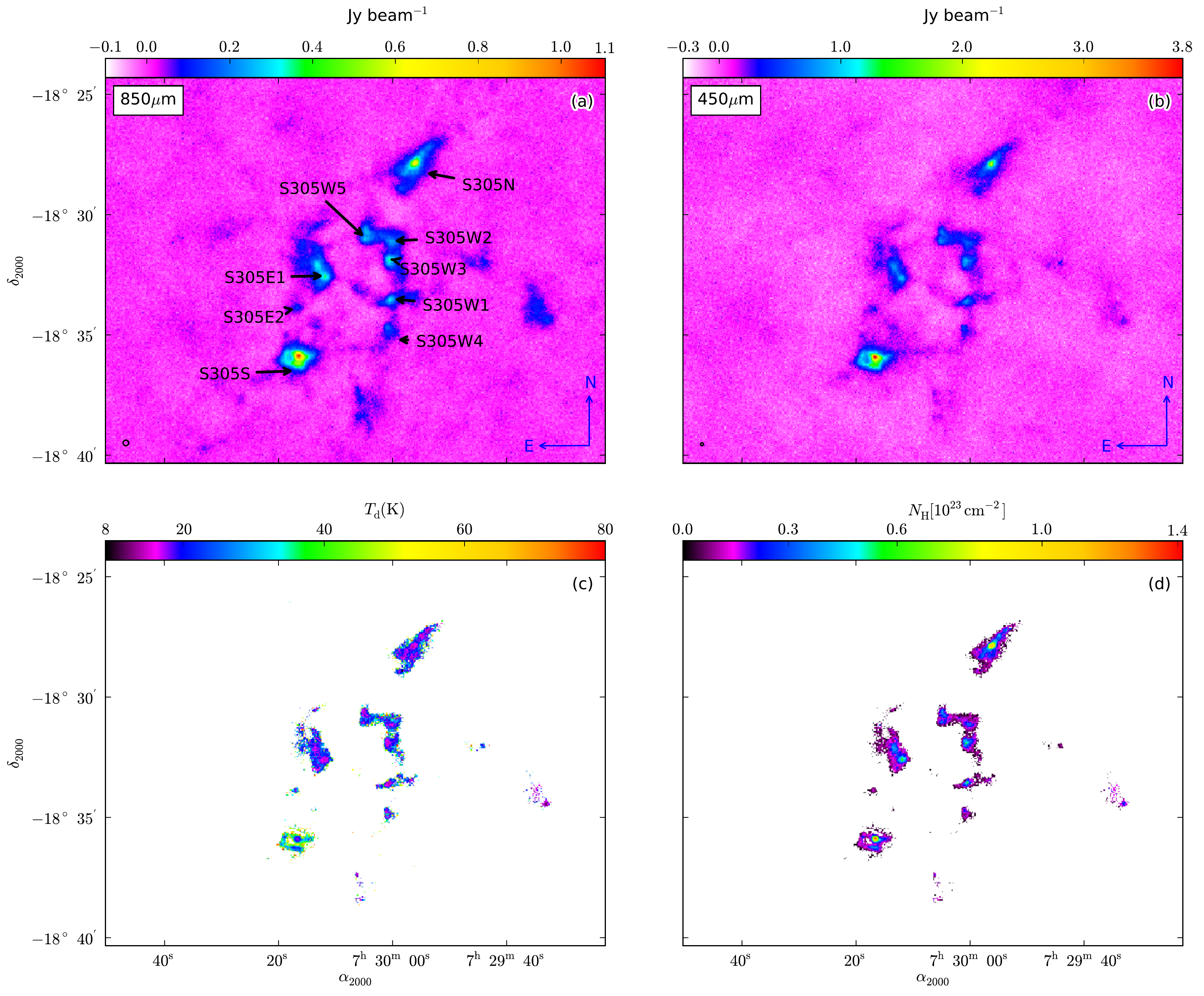} 
\end{center}
\caption{S305: (a) surface brightness map at 850 $\mu$m; (b) surface brightness map at 450 $\mu$m; (c) dust temperature map; and (d) column density map.}
\label{S305_panel}
\end{figure}

The maps of the hydrogen column density $N_{\rm{H}}$ derived from the 850 $\mu$m optical depth maps are shown in Figures \ref{S148_panel}(d), \ref{S156_panel}(d), \ref{S159_panel}(d), \ref{S254_panel}(d), and \ref{S305_panel}(d). The value of $N_{\rm{H}}$ at the outer cloud boundary is very low, $\approx$ 10$^{21}$ cm$^{-2}$ (S148, S156, S159, S305) and $\approx$ 10$^{22}$ cm$^{-2}$ (central regions of S254 complex). The column densities approach their highest values towards the centres of the clumps.  Note that a column density of $\approx$ 10$^{21}$ cm$^{-2}$ corresponds to an extinction of less than one magnitude, and thus these observed edges of the clouds should match very well with the cloud edges that one would find in an optical image. In fact, in the 450 and especially in the 850 $\mu$m images we do see slightly further out than we can accurately compute the temperature and column density, because of low S/N, reinforcing the suggestion that we are seeing the entire cloud.

The innermost region of BFS15 displays the highest column density, which is 5.3 $\times$10$^{23}$ cm$^{-2}$. Note that S254S1 is also comparably dense, with a column density of 4.3 $\times$10$^{23}$ cm$^{-2}$. Column density maps of the other clumps such as S254N, S159, S156 and the clump S305S associated with the \HII\ region S305, also demonstrate peak column densities higher than 10$^{23}$ cm$^{-2}$ at the centres. Visual extinction values of the complexes are obtained from their derived column density maps. All the clumps, except those near S148 and S305, show peak visual extinction values above 100. The highest peak $A_{V}$ (283) is found in the clump near BFS15. In S254N, S254S1, and S254S2, the peak $A_{V}$ values are, respectively, 199, 229 and 41. The other clumps in the sample (near \HII\ regions S159, S305 and S148) are characterized by peak $A_{V}$ between 63 and 198. The mean central number density, $\overline{n_{\rm{c}}}$ of each clump, is the average number density along the line of sight, obtained from the peak column density divided by the size of each clump. The mean central density of the clumps is typically about 8$\times$10$^{4}$ cm$^{-3}$. However, the central densities of the clumps range over an order of magnitude, from 10$^{4}$ to 10$^{5}$ cm$^{-3}$. The value of $\overline{n_{\rm{c}}}$ is lower than the true central density, as it is an average that includes the lower density outer parts of the clump along the line of sight. 

The derived clump properties are listed in Table \ref{derived}. The first column gives the source names. The second, third, and fourth columns list the peak values of the optical depth, the column density, and visual extinction at the centres of the clumps using the 850 $\mu$m data. The size of the clump at 850 $\mu$m and the average central number density values ($\overline{n_{c}}$) are given in the fifth and sixth columns, respectively. The last column provide the values of the central dust temperatures of the clumps. Uncertainties in the diameter and number densities of the clumps are about 18\% (due to uncertainties in the cloud distances) and not given in the table. Additional calibration errors of 10\% at 450 and 5\% at 850 $\mu$m in the SCUBA-2 flux densities also contribute 11\% uncertainties to the values of $\tau_{850}$, N$_{\rm{H}}$, A$_{V}$, $\overline{n_{\rm{c}}}$, and ${{T_{\rm{c}}}_{\rm{d}}}$ (not given in the Table). These calculations were repeated for $\beta=1.5$ (see Table \ref{derived_second}) and for $\beta=2.5$ (see Table \ref{derived_third}).

\begin{table}[ht]
\caption{Properties of the clumps: Column 1 lists the clump names associated with the Galactic \HII\ regions. Columns 2, 3, and 4 list the peak values of the optical depth, column density, and visual extinction (using a $\beta$ of 2) detected towards the centres of the respective clumps. The The 850 $\mu$m size, the central number densities, and the central dust temperatures (for $\beta$= 2) of the clumps are listed in the fifth, sixth, and seventh columns, respectively.}
\label{derived}
\begin{center}
\leavevmode
\begin{tabular}{l c c c c c l} \hline \hline 
            
Source & $\tau_{850}$ &$N_{\rm{H}}$ & $A_{V}$  & Diameter &$\overline{n_{\rm{c}}}$&${{T_{\rm{c}}}_{\rm{d}}}$ \\ 
	      && (10$^{23}$ cm$^{-2}$)& (mag)& (pc) &(cm$^{-3}$) & (K)\\
\hline
S148\tablenotemark{1}	&6.3$\times$10$^{-3}$		&2.0		&86.1	&3.56	&1.5$\times$10$^{4}$	&\nodata	\\ 
S156			&9.9$\times$10$^{-3}$		&2.6		&136		&3.06	&6.1$\times$10$^{4}$	&11.5	\\
BFS15		&0.02					&5.3		&283		&1.80	&1.6$\times$10$^{5}$	&9.5		\\
S159			&0.01					&3.7		&198		&1.62	&7.4$\times$10$^{4}$	&14.3	\\
S254N		&0.01					&3.7		&199		&0.76	&1.3$\times$10$^{5}$	&24.6	\\
S254S1		&0.02					&4.3		&229		&0.90	&1.8$\times$10$^{5}$	&23.4	\\
S254S2		&3.0$\times$10$^{-3}$		&0.8		&41		&0.51	&4.9$\times$10$^{4}$	&28.6	\\
S305N		&4.6$\times$10$^{-3}$		&1.2		&63		&1.62	&2.3$\times$10$^{4}$	&13.7	\\
S305S		&5.4$\times$10$^{-3}$		&1.4		&73		&2.31	&2.0$\times$10$^{4}$	&16.7	\\
\hline
\end{tabular}
\tablenotetext{1}{By assuming a constant temperature of 10 K throughout the cloud.}
\end{center}
\end{table}

\begin{table}[ht]
\caption{ Similar to Table \ref{derived}, but using $\beta=1.5$. Column 1 lists the clump names associated with the Galactic \HII\ regions. Columns 2, 3, and 4 list the peak values of the optical depth, column density, and visual extinction detected towards the centres of the respective clumps. The central number densities and central dust temperatures of the clumps are listed in the fifth and sixth columns, respectively.}
\label{derived_second}
\begin{center}
\leavevmode
\begin{tabular}{l c c c c l} \hline \hline 
            
Source  &$\tau_{850}$ &N$_{\rm{H}}$ & A$_{V}$  &$\overline{n_{\rm{c}}}$ & ${{T_{\rm{c}}}_{\rm{d}}}$ \\ 
	    &  & (10$^{23}$ cm$^{-2}$)& (mag)&(cm$^{-3}$)  & (K) \\
\hline
S148					&\nodata							&\nodata		&\nodata		&\nodata					&\nodata		\\ 
S156		 			&5.1$\times$10$^{-3}$				&1.3			&70			&3.1$\times$10$^{4}$		& 	16.9		\\
BFS15				&0.01							&2.9			&154			&8.9$\times$10$^{4}$		&	12.7		\\
S159					&6.2$\times$10$^{-3}$				&1.6			&85			&3.2$\times$10$^{4}$		&	24.9		\\
S254N				&\nodata							&\nodata		&\nodata		&\nodata					&\nodata		\\
S254S1				&\nodata							&\nodata		&\nodata		&\nodata					&\nodata		\\
S254S2 				&\nodata							&\nodata		&\nodata		&\nodata					&\nodata		\\
S305N				&2.0$\times$10$^{-3}$				&0.52		&28			&1.0$\times$10$^{4}$		&	23.1		\\
S305S				&1.9$\times$10$^{-3}$				&0.48		&26			&6.8$\times$10$^{3}$		&	35.1		\\			
\hline
\end{tabular}
\end{center}
\end{table}

\begin{table}[ht]
\caption{ Similar to Table \ref{derived}, but using $\beta=2.5$. Column 1 lists the clump names associated with the Galactic \HII\ regions. Columns 2, 3, and 4 list the peak values of the optical depth, column density, and visual extinction detected towards the centres of the respective clumps. The central number densities and central dust temperatures of the clumps are listed in the fifth and sixth columns, respectively.}
\label{derived_third}
\begin{center}
\leavevmode
\begin{tabular}{l c c c c l} \hline \hline 
            
Source  & $\tau_{850}$ &N$_{\rm{H}}$ & A$_{V}$  &$\overline{n_{\rm{c}}}$ & ${{T_{\rm{c}}}_{\rm{d}}}$ \\ 
	    &  & (10$^{23}$ cm$^{-2}$)& (mag)&(cm$^{-3}$) &(K) \\
\hline
S148		      			& \nodata 				&\nodata				&\nodata	&\nodata				&\nodata	\\ 
S156		 	 		&0.02				&4.3					&232		& 1.0	$\times$10$^{5}$	& 8.9	\\
BFS15	 			&0.03				&8.8					&469		&  2.7$\times$10$^{5}$	& 7.7	\\
S159					&0.03				&6.7					&360		& 1.3$\times$10$^{5}$	&10.3\\
S254N			 	&0.03				&8.7					&462		& 3.1$\times$10$^{5}$	&14.2\\
S254S1				&0.04				&9.7					&518		& 4.2$\times$10$^{5}$	&13.8\\
S254S2				&7.6$\times$10$^{-3}$	&1.9					&104		& 1.2	$\times$10$^{5}$	&15.3\\
S305N				&8.2$\times$10$^{-3}$	&2.1					&113		& 4.2$\times$10$^{4}$	&10.1\\
S305S				&0.01				&2.7					&142		& 3.7$\times$10$^{4}$	&11.4\\			
\hline
\end{tabular}
\end{center}
\end{table}

\subsubsection{Gas Masses from Dust Emission}

The gas masses are derived by summing up all the pixel values from the column density maps of the clouds, which are generated from the 850 $\mu$m dust emission maps, employing a dust temperature distribution assuming a value of $\beta$. i.e., 
$
M_{\rm{gas}} = \mu m_{\rm{H}} N_{\rm{H}} A  
$, 
where $N_{\rm{H}}$ is the total hydrogen column density in cm$^{-2}$, and $A$ is the pixel area in cm$^{2}$. The gas masses of the clouds are listed in Table \ref{mass}. The first column lists the cloud names. The second, third, and fourth columns list the gas masses corresponding to $\beta$ being 2, 1.5, and 2.5, respectively. The uncertainties in the masses are due to the cloud's shape and this far exceed the flux uncertainties. The mass uncertainties do not include the calibration uncertainties. Flux calibration uncertainties are $\sim$10\% at 450 and 5\% at 850 $\mu$m. The letter ``C'' associated with a cloud in Table \ref{mass} refers to a cloud complex: masses of the S159, S305, and S254 complexes are also included in the Table. All the individual clouds in the complexes are massive with values of gas mass ranging from 11 M$_{\odot}$ to 9.9$\times$10$^{2}$ M$_{\odot}$. The least massive cloud is S156NE, with a gas mass of 11 M$_{\odot}$, while the most massive one is S305 complex (S305C) with a total mass of 3.1$\times$10$^{3}$ M$_{\odot}$, at least for the $\beta$=2 case. When $\beta$=2.5, the least massive cloud is S305E2 (39 M$_{\odot}$) and the most most massive one is S305C (3.1$\times$10$^{3}$ M$_{\odot}$). However, when $\beta$=1.5, we cannot  derive temperature maps of S254C, S156NE, and S305E2 and therefore masses of these clouds are not derived.  The difference in mass in all the three cases is because of different cloud temperatures corresponding to different values of $\beta$. 

\begin{table}[ht]
\caption{Cloud masses}
\label{mass}
\begin{center}
\leavevmode
\begin{tabular}{l c c c} \hline \hline 
            
Source  &$M_{\rm{gas}}$  (M$_{\odot}$) & $M_{\rm{gas}}$  (M$_{\odot}$) &$M_{\rm{gas}}$  (M$_{\odot}$)							\\ 
	     & $\beta=2$  &   $\beta=1.5$  &$\beta=2.5$																		\\
\hline
S148\tablenotemark{1}&(7.3$\pm$1.3)$\times$10$^{3}$& \nodata					&		\nodata						\\ 
\\
S156			&(9.1$\pm$1.0)$\times$10$^{2}$  	&(3.1$\pm$0.3$\times$10$^{2}$	& 	(1.7$\pm$0.4)$\times$10$^{3}$			\\
BFS15		&(9.9$\pm$0.6)$\times$10$^{2}$   	&(4.0$\pm$0.2$\times$10$^{2}$	& 	(1.6$\pm$0.2)$\times$10$^{3}$ 		\\
S156NE		&11$\pm$4 					& \nodata						&	 (0.6$\pm$0.2)$\times$10$^{2}$		\\
\\		
S159			&(8.9$\pm$0.6)$\times$10$^{2}$ 	&(1.0$\pm$0.4)$\times$10$^{2}$	& 	(2.3$\pm$0.1)$\times$10$^{3}$ 		\\
S159C		&(1.1$\pm$0.1)$\times$10$^{3}$ 	&(1.7$\pm$0.5)$\times$10$^{2}$ 	& 	(2.6$\pm$0.1)$\times$10$^{3}$			\\
\\
S254N		&(1.3$\pm$0.4)$\times$10$^{2}$	&\nodata						& 	(6.6$\pm$1.9)$\times$10$^{2}$			\\
S254S1		&(1.1$\pm$0.3)$\times$10$^{2}$	&\nodata						& 	(6.0$\pm$1.2)$\times$10$^{2}$			\\
S254S2		&19$\pm$7					&\nodata						&	(1.3$\pm$0.6)$\times$10$^{2}$			\\
S254N\&S		&(2.8$\pm$0.5)$\times$10$^{2}$	&\nodata						&   	(1.5$\pm$0.2)$\times$10$^{3}$			\\
S254C		&(4.6$\pm$0.7)$\times$10$^{2}$	&\nodata						&	(1.9$\pm$0.3)$\times$10$^{3}$			\\
\\
S305N		&(8.9$\pm$4.0)$\times$10$^{2}$	&(2.3$\pm$1.0)$\times$10$^{2}$ 	& 	(1.8$\pm$0.9)$\times$10$^{3}$			\\
S305W1		&(1.7$\pm$0.6)$\times$10$^{2}$	& 48$\pm$7					&	(3.8$\pm$1.5)$\times$10$^{2}$			\\
S305W2		&(2.0$\pm$0.9)$\times$10$^{2}$	& 41$\pm$9					&	(4.5$\pm$2.4)$\times$10$^{2}$			\\
S305W3		&(2.6$\pm$1.1)$\times$10$^{2}$	& 71$\pm$17					&	(5.1$\pm$2.7)$\times$10$^{2}$			\\
S305W4 		&(0.7$\pm$0.2)$\times$10$^{2}$	&$\leq$18						&	(1.5$\pm$0.6)$\times$10$^{2}$			\\
S305W5		&(2.4$\pm$0.6)$\times$10$^{2}$	&57$\pm$8					&	(5.4$\pm$1.5)$\times$10$^{2}$			\\
S305E1		&(7.3$\pm$2.3)$\times$10$^{2}$ 	&(1.6$\pm$0.4)$\times$10$^{2}$	& 	(1.5$\pm$0.6)$\times$10$^{3}$ 		\\
S305S		&(5.6$\pm$2.1)$\times$10$^{2}$ 	&$\leq$37						& 	(1.7$\pm$0.6)$\times$10$^{3}$ 		\\	
S305E2		&15$\pm$4					&\nodata						&	39$\pm$14						\\
S305C		&(3.1$\pm$0.5)$\times$10$^{3}$	&$\leq$6.6$\times$10$^{2}$		&     	(7.1$\pm$1.3)$\times$10$^{3}$			\\

\hline
\end{tabular}
\tablenotetext{1}{By assuming a constant temperature of 10 K throughout the cloud.}
\end{center}
\end{table}

\subsection{Analysis of Radial Intensity Profiles}
Predictions of different star formation scenarios are dependent on the true density distribution of the clumps. In order to investigate this, we construct and analyse intensity profiles of the clumps and discuss their implications for the star formation process. Without doing extensive radiative transfer modelling it is difficult to analyze the surface brightness profiles in any detail. Simple models using power-law density and temperature distributions have been used by \citet{Shirley00}. We follow their analysis technique but caution that our sample are significantly different from the \citet{Shirley00} objects. Their objects appear to have embedded protostars. Our sample do not have obvious embedded protostars and are cold at their centres and thus it is likely that they do not have any central heating sources.

\begin{figure}[ht!]
\begin{center}
\includegraphics[scale=0.65]{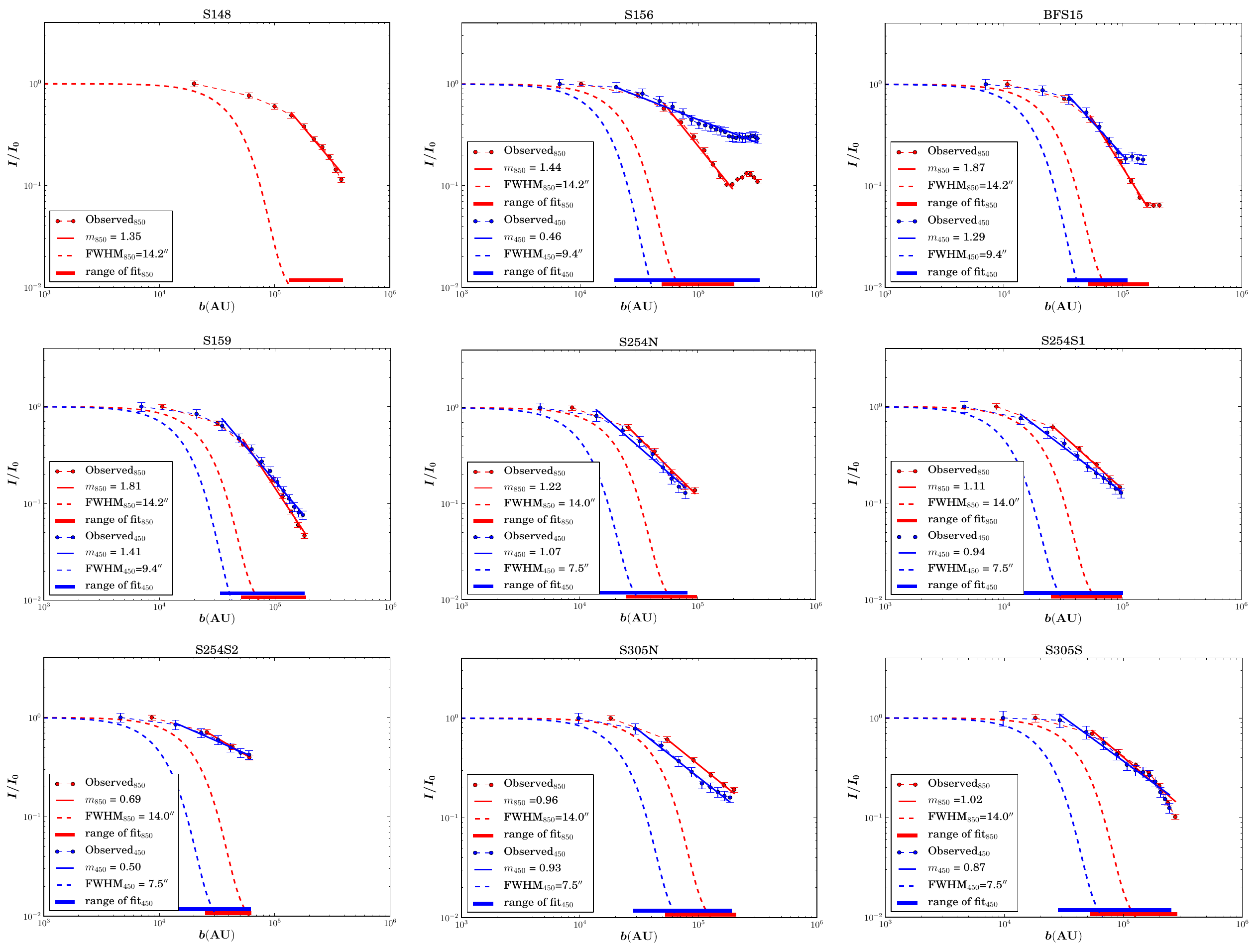}
\end{center}
\caption{Radial profiles of the sources: the 850 (in red) and 450 $\mu$m (in blue) normalized intensity profiles of the sources plotted as a function of the impact parameter $b$ (AU) along with the Gaussian main beam profile at 850 (red dashed curve) and 450 $\mu$m (blue dashed curve) wavelengths. The thick solid red and blue lines represent the range of the 850 and 450 $\mu$m fits respectively.}
\label{int_prof}
\end{figure}

Normalized, azimuthally-averaged radial profiles were constructed for each intensity map at both SCUBA-2 wavelengths. The images were binned at half the beam width to create equally-spaced annuli from the peak intensity of the clumps. Uncertainties were estimated from the quadratic sum of the errors due to calibration and the deviation from the azimuthal symmetry at each radial annulus. We calculate the mean value of the intensity $I_{\nu}(b)$ normalized to the peak intensity $I_{\nu}(0)$ at each radial bin about the impact parameter $b$ (in AU). Following the technique described in \citet{Shirley00}, power-law fits were performed for all clumps using the form $I_{\nu}(b)/I_{\nu}(0)$= $[b/b(0)]^{-m}$, with $b(0)$ corresponding to one quarter of the beam width. As in \citet{Shirley00}, we only fit each radial profile outside of the FWHM of the beam in each case. The intensity profiles in almost all cases are well-fit by a power law [see Figure \ref{int_prof}]. We find a range of values for the power law indices with the average value of 1.27$\pm$0.37 at 850 $\mu$m and 0.93$\pm$0.31 at 450 $\mu$m.

The SCUBA-2 beams are much larger than the inner radii of the radial profiles shown in Figure \ref{int_prof}. Typical physical resolution is greater than $\sim$ 0.1pc, or $\sim$ 2$\times$10$^4$ AU, which is the 450 $\mu$m SCUBA-2 beam width of 7.5$\arcsec$ at a distance of 2.46 kpc---the distance to the closest object in our sample. The average value of $m$, obtained by combining the slopes at 850 and 450 $\mu$m wavelengths, is 1.10$\pm$0.34. This value agrees well with the outer slope of 1 in the intensity distribution of isolated low mass pre-protostellar cores studied by \citet{Ward94}. Our average value of $m$ is slightly flatter than the mean slope of 1.48$\pm$0.35 obtained for low mass Class 0/I sources by \citet{Shirley00} and is consistent with the mean inner power-law intensity index of 1.2 found in massive star forming regions by \citet{Beuther02}. 

If we consider an optically thin uniform opacity spherically symmetric clump with $I_{\nu}(b)$ $\propto$ $b^{-m}$ and dust temperature $T_{\rm{d}}(r)$ $\propto$ $r^{-q}$, then the density profile of the clump $\rho(r)$  will follow $r^{-p}$, where the power-law indices can be related as $m=p+q-1$ \citep{Adams91}. The slopes of the radial temperature profiles (corresponding to $\beta= 2$) of our objects, calculated from the dust temperature maps, are used to to calculate the density power law index $p$. Since we found an average $q$ of $-$0.26$\pm$0.10 in five of our clumps, the average value of $m$ therefore translates into an average $p$ of 2.36$\pm$0.35. All the uncertainty values represent uncertainties due to the standard deviation of the sample.

The mean value of $p$ from our study is consistent with the mean value of the density power law index ($p$=2.15$\pm$0.35) corresponding to a $\beta$ of 2 in low mass Class 0/I sources \citep{Shirley00}. However, the cores in the study of \citet{Shirley00} are centrally heated, whereas all the clumps in our study have decreasing temperatures toward the centres. Our value of $p$ is steeper than the density power law index, obtained using one dimensional (1-D) radiative transfer codes, of 1.8$\pm$0.4 [\citep{Mueller02}---in dense cores associated with massive star-forming regions] and 1.3$\pm$ 0.4 [\citep{williams05}---in a sample of high-mass protostellar objects]. The $p$ values obtained using power law fits such as 1.6$\pm$0.5 [\citep{Beuther02}---in a large sample of massive cores in the earliest stages of evolution] and 1.6$\pm$0.3 [\citep{Pirogov09}---in high mass star-forming regions of the southern hemisphere] are also shallower than those from our study. Similarly, 1-D radiative transfer studies by \citet{Jorgensen02} on low-mass Class 0 and Class 1 sources show flatter density power law indices in the range 1.3 to 1.9 ($\pm$0.2) within $\sim$ 10$^{4}$ AU. However, all these values are dependent on the assumed values of the dust emissivity index, $\beta$.

Our power law indices for densities are steeper than all of these results found by other authors. However, all of these other results are for cores with embedded sources while our objects do not show any signs of embedded sources. Perhaps this difference in indices is due to this difference in the evolutionary state of the objects. Hydrostatic and hydrodynamic configurations of pressure-bounded cores have density profiles in their outer envelopes that fall off like $r^{-2}$ similar to what we have found in our studies. The study \citep{Tafalla02} of starless cores in low mass star forming regions reveals power law indices, ranging from 2 to 4 at large radii, consistent with our mean value of $p$. If our objects are truly starless, then we would expect the density structure to flatten in the inner regions. Nevertheless, our resolution (and the source distance) severely limits the ability to see this, and all the power-law structure observed in our study is outside of typically 10$^{4}$ AU.

\subsection{Caveats}\label{anal} 
The observed flux densities from sub-millimetre thermal emission are plagued by a number of uncertainties, which can influence the derived dust temperatures and densities. These uncertainties include possible line contamination, variations in the calibration factor, discrepancies in the adopted convolution method, and variations of the value of $\beta$. In addition, the assumptions used in the analysis of radial intensity profiles are sensitive to Rayleigh-Jeans limit.

Spectral line contamination in the SCUBA-2 bolometer bandpasses may be a significant source of uncertainty in the extracted physical properties. Abundant molecules such as CO and its isotopomers embedded in dust clouds are one of the primary sources of dust continuum pollutants at sub-millimetre wavelengths \citep{Gordon95, Papa00, Seaquist04}. The major contributor to line emission in the 850 micron band is expected to be the CO(3-2) line at 345 GHz. Methanol and SO$_{2}$ are likely to be significant contributors to the continuum emission---although much weaker than the CO lines, these molecules emit in dozens or even hundreds of separate lines within the bandpass. We can estimate the effect on our results if there is a significant amount of line emission  in our 850 $\mu$m images but none in our 450 $\mu$m images. This is effectively a ``worst'' case, because the presence of line emission in the 850 $\mu$m image would cause us to infer a lower temperature, while such emission in the 450 $\mu$m would move our calculated temperatures higher. However, it is possible that in our clouds the line emission is very weak, because the regions are very cold. For example, the CO emission in the 850 $\mu$m band is the J=3$\rightarrow$2 transition, which is bright in warm regions. We have CO(3-2) data for some of our objects, and as we discuss below, we find this is not an important effect in the denser, colder centres of these clouds. The CO has a significant effect on the observed fluxes in the hot outer edges of most of these clouds.

We have obtained CO(3-2) data from the JCMT archives\footnote{http://www.cadc-ccda.hia-iha.nrc-cnrc.gc.ca/jcmt/} for four of our objects: S156, S159, S254, and S305S. The CO(3-2) contribution to the SCUBA-2 850 $\mu$m dust continuum emission is found to be below $\sim$ 2-3\% toward the centres of S159, S254N, and S254S, and $\sim$10\% at the outermost borders of the clumps. The extended envelopes of the S159 cloud exhibit higher contributions from CO(3-2) line: $\sim$15-45\% for the north-west and east, $\sim$40-90\% for the north-east envelope. A few pixels at the north-west part of the central S254 cloud show dominant CO(3-2) line contribution, $\sim$30-60\%. Overall, CO(3-2) line contributes little to the bulk of the S254 cloud. Contamination under $\sim$15\% towards the centres and $<$40\% towards the outer borders of the clumps S156 and S305S are observed. The CO(3-2) line emission contributed $\sim$60 to 90\% of the SCUBA-2 850 $\mu$m flux at some parts of the the extended envelope of S156. Recent studies on Perseus B1 clump by \citet{Sadavoy13} found that CO(3-2) line emission contributed only less than 15\% of the 850 $\mu$m flux on the internally heated regions and less than 1\% of the 850 $\mu$m flux on the colder regions of the B1, with the most significant contribution, $\sim$ 90\%, at the outflow positions. Studies on a number of nebulae in the Perseus and Orion cloud complexes by \citet{Drabek12} confirmed less than 20\% contribution from the CO(3-2) line emission on the SCUBA-2 850 $\mu$m flux in regions without outflows and an insignificant contribution from CO(6-5) lines at 450 $\mu$m dust emission.

\begin{figure}[ht]
\begin{center}
\includegraphics[scale=0.66]{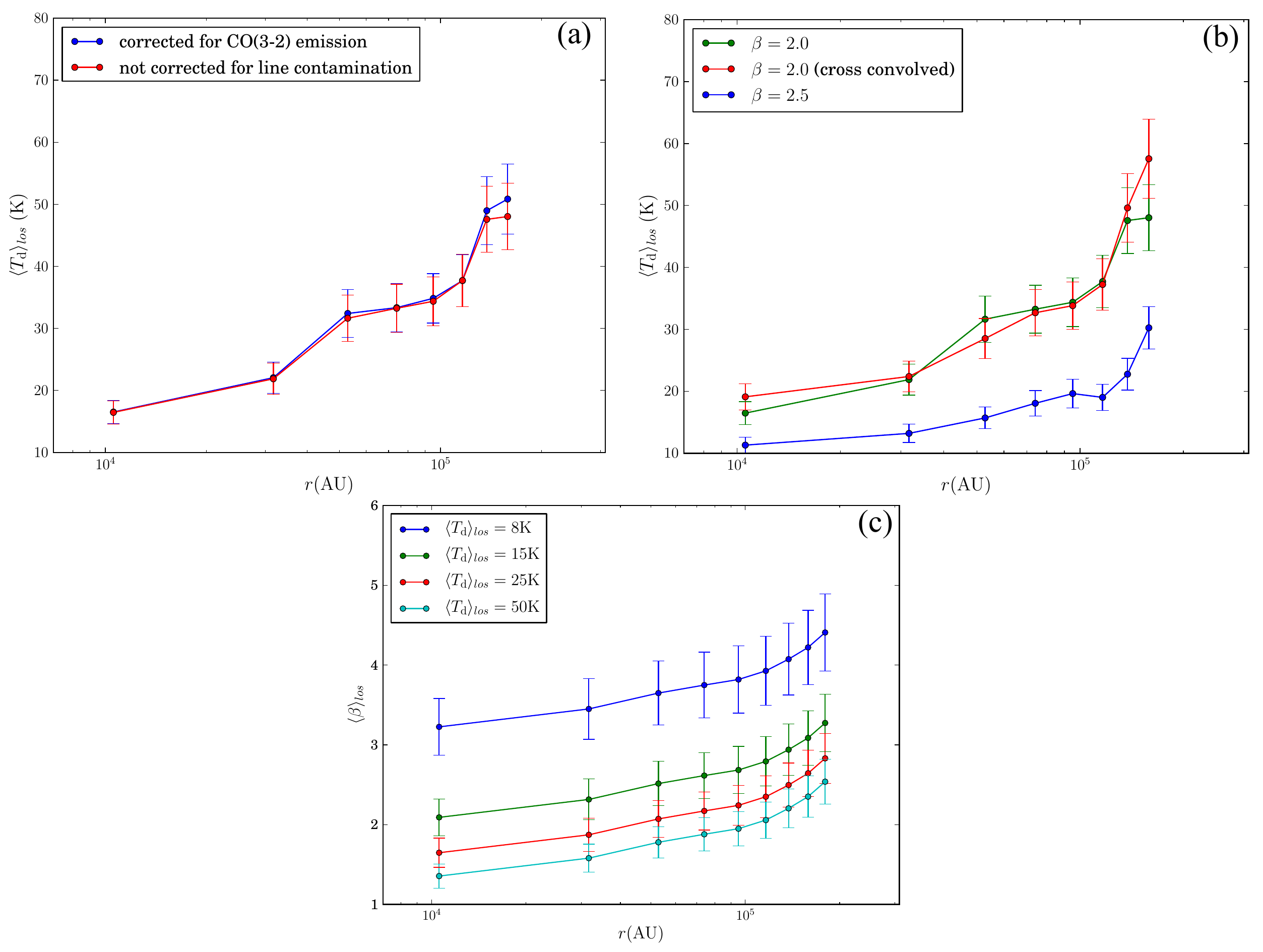}
\end{center}
\caption{(a) Line-of-sight averaged dust temperature ( $\langle T_{\rm{d}} \rangle_{los}$) profiles of the cloud associated with S159, with (red curve) and without (blue curve) line contamination at 850 $\mu$m. Blue curve represents temperature obtained from the data corrected for CO(3-2) emission at 850 $\mu$m, using a $\beta$ of 2. (b) Variations in  $\langle T_{\rm{d}} \rangle_{los}$ profile with different assumed values of $\beta$. The red curve represents the temperature obtained from the 450/850 ratio map, using a $\beta$ of 2, after doing cross-convolution [convolving 450 $\mu$m data to the resolution of 850 $\mu$m 14.2$\arcsec$ primary beam and 850 $\mu$m data to the resolution of 450 $\mu$m 9.4$\arcsec$ primary beam]. Error bars indicate the uncertainty in the temperature due to a combination of calibration and the error in mean of the temperature distribution in each radial annulus. (c) Variations in $\beta$ corresponding to different possible values of $\langle T_{\rm{d}} \rangle_{los}$ from 8 K to 50 K. Error bars indicate the uncertainty in $\beta$ due to a combination of calibration and the error in mean of the $\beta$ distribution in each radial annulus.}
\label{S159_Tdr_cont_multibeta}
\end{figure}

The removal of line-contaminated flux densities from the SCUBA-2 850 $\mu$m data has only a minor effect on the dust temperatures in the dense clumps [see Figure \ref{S159_Tdr_cont_multibeta}(a)]: subtraction of the observed S159 CO(3-2) data from the 850 $\mu$m SCUBA-2 data barely changed the central dust temperatures although it slightly increased the outer temperatures [Figure \ref{S159_Tdr_cont_multibeta}(a)--blue curve]. If there is a higher contribution from CO or from other molecular lines within this band (for example $\sim$30\%), then the clump temperatures alter considerably: removal of an assumed 30\% line contamination increased the temperatures by around 93\% at the centre. However, due to very high 450/850 flux ratio at the edges, the temperatures at the clump edges could not be evaluated. Elimination of the line contribution to the continuum (using the observed CO data) did not change the density power law index. Although the CO(3-2) emission does not affect the main results of this paper (very cold, dense clumps), it does cause us to slightly under-estimate the temperature at the edges of the clouds surrounding the clumps. A more detailed analysis of the CO emission is underway and will appear in a future paper.

Small variations in the adopted value of flux calibration factor have a negligible effect on the results. For example, employing the standard value of FCF to calibrate the S156 map field, by raising the value of FCF by $\sim$2\% at 850 and $\sim$7\% at 450, caused only $\sim$4-5\% variations in the central dust temperatures and $\sim$ 7\% variations in the peak dust column densities of the clouds.

Since the 850 $\mu$m beam solid angle is larger due to the error beam pattern than just the main beam solid angle and there is large difference in the side-lobe patterns between 450 and 850 $\mu$m, convolving 450 $\mu$m SCUBA-2 map to the 850 $\mu$m resolution of the JCMT beam, to roughly match the Gaussian main beams at both the wavelengths, may lead to some uncertainties in the result. We, therefore, performed a cross-convolution on one of the cloud (S159), by convolving 450 $\mu$m data to the resolution of 850 $\mu$m 14.2$\arcsec$ primary beam and 850 $\mu$m data to the resolution of 450 $\mu$m 9.4$\arcsec$ primary beam as described in \citet{Hatchell13}. The cross convolution method raised the value of the dust temperature at the centre of S159 by around 3 K. However, this method did not make a large increase in the temperature and all the values obtained from this method are within the uncertainty limits of the temperatures obtained by the method used in this paper [see Figure \ref{S159_Tdr_cont_multibeta} (b)].

Variations in the $\langle T_{\rm{d}} \rangle_{los}$ profiles of the clump close to S159, corresponding to different values of $\beta$ are displayed in Figure \ref{S159_Tdr_cont_multibeta}(b). Since there is some evidence of $\beta$ $>$2 toward low-mass star forming regions \citep{Schnee10,Shirley11}, we also explored the possibility of $\beta$ $=$ 2.5 in our calculations.  As the value of $\beta$ increases from 2.0 to 2.5, the $\langle T_{\rm{d}} \rangle_{los}$ profile shifts downwards denoting an anti-correlation between temperature and values of $\beta$, a very well-known effect, discussed by many authors. Our determination of the temperatures depends on what value of $\beta$ we use in the calculations. The central clump temperatures shift from $\sim$ 10 K to 14 K as we decrease the value of $\beta$ from 2.5 to 2.0, indicating cold central regions even in the presence of lower $\beta$. Similarly, between $\beta$ of 1.5 and 2.5, \citet{Sadavoy13} found $\sim$ 4 K variation in the dust temperature towards B1-a core. However, when $\beta$ is 1.5, the central clump temperatures of S159 is $\sim$ 25 K but we are unable to plot a temperature profile due to high values of 450/850 ratio toward the outer layers. The change in $\beta$ barely affects the density power-law index; for example, as we increase the value of $\beta$ from 2 to 2.5, the average value of $p$ increases by merely 1\%. The clouds show another conspicuous feature: regions of lower temperatures correspond to higher column densities if $\beta$ is uniform throughout the cloud. This feature might imply that the dust grains in the densest regions of the clouds are effectively shielded from the stellar radiation field. As a consequence, grains in these regions are more likely to coalesce than shatter due to lower average kinetic energy \citep{Chokshi93}.

Using the emission from the two SCUBA-2 wavelengths, we computed the value of the dust emissivity index in each pixel by assuming a constant line of sight averaged dust temperature of 8 K, 15 K, 25 K, and 50 K throughout the cloud [see Figure \ref{S159_Tdr_cont_multibeta}(c)]. The assumed temperatures  are considered to be typical temperature that can be found in different radial layers of the cold giant molecular clouds. A radial variation of $\beta$ is observed in all the complexes at all selected temperatures, with low values (for e.g., $\beta$ $\sim$1.3 at $\langle T_{\rm{d}} \rangle_{los}$=50 K) and high values ($\beta \approx 3.2$ at $\langle T_{\rm{d}} \rangle_{los}$=8 K) towards the centre of S159. Increasing the assumed temperature from 8 K to 50 K produces on average a decrease in $\beta$ by 59\% toward the centre of S159. Much lower values of $\beta$ ($\sim0.5$) are observed toward the centre of the south-west fragment cloud in the S159 complex when $\langle T_{\rm{d}} \rangle_{los}$=50 K. Note that the lowest value of $\beta$ detected in the direction of the isolated low-mass star-forming core TMC-1C's centre \citep{Schnee05} is  $\sim$ 1.7. At each $\langle T_{\rm{d}} \rangle_{los}$, one can see lower values of $\langle \beta \rangle_{los}$ towards the centres of the clouds. Calculating variations in $\langle T_{\rm{d}} \rangle_{los}$ by assuming $\langle \beta \rangle_{los}$  $\sim$ constant is probably much closer to reality than calculating variations in $\langle \beta \rangle_{los}$ by assuming $\langle T_{\rm{d}} \rangle_{los}$ is constant because the range of $\langle \beta \rangle_{los}$ in the latter case is far outside anything actually measured at sub-millimetre regime (e.g., up to 3.2 in the centres).

The assumption that $m=p+q-1$, used in the analysis of radial intensity profiles of the clumps, is valid only in the Rayleigh-Jeans limit, where ($h \nu /k$)$<<T$. At temperatures below 17 K at 850 microns, the Rayleigh Jeans law breaks down and the interiors of our clumps, wherever ${T_{\rm{d}}}$ is less than 17 K, will not satisfy power law equations. To correct for this, one needs to go back to the \citet{Adams91} analysis and then change it to account for the low temperatures. However, this will not allow us to directly compare the results to \citet{Shirley00}, \citet{Beuther02}, and \citet{Pirogov09}, as we have done in this paper. Furthermore, modifying the expressions of \citet{Adams91} is mathematically too challenging for the current paper.Ê

\section{DISCUSSION}\label{dis}
Stars usually form in locations of cold, dense molecular gas at high densities, over 10$^{4}$ cm$^{-3}$, and giant molecular clouds (GMCs) are ideal sites for such dense concentrations of gas. However, the star formation efficiency, defined as the star formation rate per unit mass of gas, in GMCs is found to be very low, less than 3\% \citep{Myers86, Lada87}, compared to the star formation efficiency in low-mass star-forming regions. The efficiency of isolated star formation is 9--15\% \citep{Swift08} and as high as 30--50\% \citep{Matzner00} for clustered low-mass star formation. However, in regions of proto-stellar outflows, the low-mass star formation efficiency can be 25--70\% \citep{Matzner00}. Analyzing the clumps within GMCs is therefore important to better understand the low efficiency of their star formation.

The temperatures of the clouds determine the initial stages of star formation. To verify the conditions in the clumps, we calculated the Jeans mass of the clumps (for $\beta$=1.5, 2, and 2.5) using central average number densities, assuming that the gas and the dust temperatures are coupled inside the clumps. At regions of high density, higher than $\approx$10$^{4}$ cm$^{-3}$, the gas and the dust temperatures share the same value \citep{Galli02, Goldsmith01}. The derived Jeans masses of the clumps are significantly smaller than the mass of the clumps calculated from the dust emission, suggesting that gravitational attraction overwhelms the thermal pressure and in the absence of other non-thermal support mechanisms, the clumps will undergo a contraction which in turn is responsible for a probable eventual collapse. Moreover, there could be cores within these clumps that are collapsing (and still starless); their existence is unknown without higher resolution observations. The support mechanisms may include turbulence or magnetic fields. Ammonia-line observations (to test for turbulence) or observations of sub-millimetre polarization (to look for ordered magnetic fields) would aid in our understanding of how these clumps survive without collapse.

We cannot constrain the values of $\beta$ and temperature simultaneously using just two wavelengths. We therefore, assumed a constant $\beta$ to calculate the temperature variations in our cloud samples. However $\beta$ may not be uniform in real clouds. Data from shorter \textit{Herschel} wavelengths (PACS 160 and SPIRE data) with SCUBA-2 data may better constrain the values of the dust temperatures because we can simultaneously calculate temperature and $\beta$. However we have not observed the clouds in \textit{Herschel} bands and therefore we calculated the physical properties of the clouds with a $\beta$ of 1.5, 2, and 2.5 to get possible range of physical values. The inclusion of filtered long wavelength SCUBA-2 data to \textit{Herschel} data has been found to increase the reliability of the temperature and $\beta$ calculation \citep{Sadavoy13}. \textit{Herschel} observations are more sensitive to larger scale structures but since such structures contribute to both 450 and 850 $\mu$m emission, it is not clear to what extent this affects our results. In any case, our focus is on the smaller scale structures---the dense clumps---and in order to get colour temperatures for these it would be necessary to remove the larger scale structure---which SCUBA-2 already does for us.

We have found cold dust, 9.5 K to 28.6 K for $\beta$ of 2 and 7.7 K to 15.3 K for  $\beta$ of 2.5, near the \HII\ regions considered in this study. But even with very shallow values of $\beta$, e.g., $\beta$=1.5 everywhere, we have found central temperatures ranging from $\sim$ 13 K to 35 K in most of the clouds. The clumps are still unstable to gravitational collapse even with such shallow values of $\beta$. All these analyses imply that our target clumps are indeed cold and that therefore there is not yet any protostar at their centres. Recent studies by \citet{Laun13} have found similar observational evidence of cold dust (8-12 K) at the interiors, with dominant heating (14-20 K) from ISRF at the outer rims, on a number of isolated low-mass starless cloud cores.

Our results of very cold dust near high-mass star forming regions are consistent with the range of theoretical minimum temperatures found in low-mass starless cores \citep{Evans01}. Besides dust continuum absorption of the ISRF, \citet{Evans01} considered different heating mechanisms that will elevate the dust temperature at the centre of a starless core/clump such as cosmic rays, effect of uv photons which are created following cosmic ray heating, and collision with warm gas. However, note that all our clouds are further from the Galactic Centre than the \citet{Evans01} clouds---in  some cases 50\% further, which means that heating by cosmic rays should be significantly less and our clouds are bigger and denser and thus heating by ISRF and UV is also considerably less, except, most likely, in the case of the central clumps in S254 complex.

In nearby, low-mass starless cores, the typical outer dust temperature is $\sim$ 20 K for ISRF intensity, $G_{0}$,  $\sim$ few. Typical outer dust temperature in our clumps is $\sim$ 60 K, then since (for $\beta$=2) outer dust temperature is $\sim$ ${G_{0}}^{(1/6)}$, it means that $G_{0}$ is $\sim$ 1000 times higher than those probed in local cores not associated with \HII\ regions. This is what we expect for clumps near strong radiation fields of O/B stars \citep{Jorgensen06}.

Not much is known about the structure of the density distribution in high mass clumps, although this is rigorously studied in low mass cores. The high mass clumps in our study possess a  steepened outer region ($\rho$ $\propto$ $r^{-2.36}$) for a $\beta$ of 2, and the power-law structure observed is outside of typically 10,000 AU. The inner radial profile of our target clumps appear to be flatter than the outer envelope, however, these regions are well within the telescope beam and are unresolved. Inner flattening of the density profiles, with $\rho$ $\propto$ $r^{-2}$ towards the outer regions, is found to be a characteristic feature of evolving low mass pre-stellar cores \citep{Ward94, Andre96, Ward99, Andre00}. An isothermal low mass spherical core that collapses to form stars follows an $r^{-2}$ density profile \citep{Shu77}. If high mass star formation is a scaled up version of low mass star formation, the cold clumps and decreasing temperatures toward the clump centres, along with steeper outer density power laws, most likely suggest that the centres have not yet gained the density required for the formation of high mass stars. These clumps are so massive that they may be fragmenting into multiple cores, but it will require interferometric observations to sort this out. High resolution observations of the clumps are therefore needed for a detailed analysis of the innermost density structure of the high mass clumps and their star formation scenarios.

The masses of the individual clouds range from around ten to a thousand solar mass, for $\beta$ of 2 and 2.5. Three of our complexes, namely S254 \citep{Heyer89}, S305, and S148 \citep{Azimlu11} were studied  using CO data. The mass of S254 complex obtained from CO observations (2.7$\times$10$^{4}$ M$_{\odot}$) is higher than the value obtained from our studies (4.6$\times$10$^{2}$ M$_{\odot}$ [$\beta$=2], 1.9$\times$10$^{3}$ M$_{\odot}$ [$\beta$=2.5]), indicating that we may have underestimated the gas mass. This probably suggests that this is a complex region because there are so many \HII\ regions. However, the mass of the S148 complex calculated from CO studies (7.9$\times$10$^{3}$ M$_{\odot}$) is close to our results (7.3$\times$10$^{3}$ M$_{\odot}$). Similarly, the mass of the S305 complex from CO studies (3.8 $\times$10$^{3}$ M$_{\odot}$) is close to the mass obtained from our studies (3.1$\times$10$^{3}$ M$_{\odot}$ [$\beta$=2], 7.1$\times$10$^{3}$ [$\beta$=2.5]).

The masses obtained from the column density maps using the observed sub-millimetre data could be an overestimation of the true cloud masses. Recent research on star-forming clumps, using simulations that model the gravitational collapse of molecular clouds, argue that the presence of lower density materials along the sightline can result in an overestimation of the cloud mass by a factor of up to three \citep{Ward12}. All our clouds are vulnerable to this effect, and there can thus be a factor of up to three decrease in the observed cloud masses. This effect is dependent on the mass: towards massive regions, this effect is more pronounced, due to the influence of much higher density material along the line of sight.

The high visual extinction, high mass, high column densities and cold temperatures of our sample clumps favour an in-falling state over an equilibrium state. At visual extinctions greater than 100 and column densities higher than 10$^{23}$ cm$^{-2}$ at very high masses, the cloud cores are unstable to collapse \citep{Kauf04}. All the observed clumps come under the category of High Mass Starless Clumps (HMSCs), which by definition do not host any massive protostars at the centre, and are the probable precursors of high-mass proto-stellar objects. HMSCs represent initial conditions of high-mass star formation and are among the least studied objects in the field of star formation. This is because these objects' peak emission falls in the sub-millimetre regime, and until recently we were unable to probe their physical properties due to the lack of sufficient instruments that operate at these wavelengths.

Very few HMSCs are known to date, and probing the properties of HMSCs is central to understanding the massive star formation process. Because these regions evolve fast, they most likely generate massive stars which live for around million years without clearing out their dust cocoon. Most of the known HMSCs \citep{Sridharan05} have masses ranging from 10${^2}$ to 10$^{3}$ M$_{\odot}$ and are at multiple evolutionary stages. Additional HMSC candidates have been identified in a handful of studies over the past couple of years. Using 870 $\mu$m ATLASGAL data, \citet{Tackenberg12} identified starless clumps with masses ranging from tens to ten thousands of solar mass, having peak column densities $>$ 10$^{23}$ cm$^{-2}$. Analysis of IRDCs using $\textit{Herschel}$ Infrared Galactic Plane Survey (Hi-Gal) data confirmed the presence of starless cores with masses ranging from 7 to 200 M$_{\odot}$ and central densities spanning $\sim$10$^{3}$ cm$^{-3}$ to $\sim$10$^{5}$ cm$^{-3}$ \citep{Wilcock12}.

Typical molecular clouds of our samples possess total IR luminosities ranging from 10$^{3}$ to 10$^{6}$ L$_{\odot}$ and some of the clouds are identified with young stellar populations and open stellar clusters---primarily the main sequence stars associated with them \citep{Azimlu10}---and are at multiple evolutionary stages. One of our SCUBA-2 targets, S254, is a multiple star-forming complex and Class I and Class II YSOs, which represent stars with infalling envelope and disk respectively, are detected in the region \citep{Fazio08}. Thus even in such complexes, we can find cold dust.

\section{CONCLUSION}
In this paper, we have studied molecular clouds near Galactic \HII\ regions, using dust continuum observations performed using the JCMT SCUBA-2 camera. A total of five Galactic \HII\ regions were targeted using SCUBA-2. The central goal of this work has been to measure the temperature and density of these clouds through the continuum emission from the dust and thereby to gain insight into the processes of massive star formation. The exterior of these clouds are subjected to intense heating from the O stars powering the \HII\ regions immediately adjacent to the clouds in our sample. However these clouds are very dense, with extremely high visual extinctions from their surfaces to most of their interior volume. The dust temperature distribution in molecular clouds near Galactic \HII\ regions was examined using the 450 and 850- $\mu$m SCUBA-2 data, leading to the identification of very dense clumps with cold dust, down to 8-17 K at the centres for emissivity indices of $\beta$ = 2.5-1.5, near hot star forming environments. All of the regions observed show signs of filamentary structure, although not as clearly in some cases as in others. We suspect that this is due to the lessened sensitivity to larger scale structures in these images. The dense clumps are found along the centres of these filaments in those cases where the filament is clearly seen. We find that these clumps are cold even after accounting for possible line contamination in the continuum and even if the dust emissivities are shallow. At these low temperatures longer wavelength observations, such as with SCUBA-2 at 850 $\mu$m have a greater ability to separate the emission from that due to hotter regions.  

\section{Acknowledgements}
The James Clerk Maxwell Telescope is operated by the Joint Astronomy Centre on behalf of the Science and Technology Facilities Council of the United Kingdom, the National Research Council of Canada, and (until 31 March 2013) the Netherlands Organisation for Scientific Research. Additional funds for the construction of SCUBA-2 were provided by the Canada Foundation for Innovation. We are grateful to Dr.\,Carolyn McCoey (University of Waterloo) for several important discussions and Dr.\,David Berry (JAC, Hawaii) for valuable assistance regarding SCUBA-2 data reduction. This research made use of \textit{pywcsgrid2}\footnote{http://leejjoon.github.com/pywcsgrid2/}---an open-source plotting package for Python.

\clearpage 
\bibliographystyle{apj}
\bibliography{SCUBA2_Dust}

\begin{thebibliography}{95}
\expandafter\ifx\csname natexlab\endcsname\relax\def\natexlab#1{#1}\fi

\bibitem[{{Adams}(1991)}]{Adams91}
{Adams}, F.~C. 1991, ApJ, 382, 544

\bibitem[{{Anderson} {et~al.}(2010){Anderson}, {Zavagno}, {Rod{\'o}n},
  {Russeil}, {Abergel}, {Ade}, {Andr{\'e}}, {Arab}, {Baluteau}, {Bernard},
  {Blagrave}, {Bontemps}, {Boulanger}, {Cohen}, {Compi{\`e}gne}, {Cox},
  {Dartois}, {Davis}, {Emery}, {Fulton}, {Gry}, {Habart}, {Huang}, {Joblin},
  {Jones}, {Kirk}, {Lagache}, {Lim}, {Madden}, {Makiwa}, {Martin},
  {Miville-Desch{\^e}nes}, {Molinari}, {Moseley}, {Motte}, {Naylor}, {Okumura},
  {Pinheiro Gon{\c c}alves}, {Polehampton}, {Saraceno}, {Sauvage}, {Sidher},
  {Spencer}, {Swinyard}, {Ward-Thompson}, \& {White}}]{Anderson10}
{Anderson}, L.~D., {Zavagno}, A., {Rod{\'o}n}, J.~A., {Russeil}, D., {Abergel},
  A., {Ade}, P., {Andr{\'e}}, P., {Arab}, H., {Baluteau}, J.-P., {Bernard},
  J.-P., {Blagrave}, K., {Bontemps}, S., {Boulanger}, F., {Cohen}, M.,
  {Compi{\`e}gne}, M., {Cox}, P., {Dartois}, E., {Davis}, G., {Emery}, R.,
  {Fulton}, T., {Gry}, C., {Habart}, E., {Huang}, M., {Joblin}, C., {Jones},
  S.~C., {Kirk}, J.~M., {Lagache}, G., {Lim}, T., {Madden}, S., {Makiwa}, G.,
  {Martin}, P., {Miville-Desch{\^e}nes}, M.-A., {Molinari}, S., {Moseley}, H.,
  {Motte}, F., {Naylor}, D.~A., {Okumura}, K., {Pinheiro Gon{\c c}alves}, D.,
  {Polehampton}, E., {Saraceno}, P., {Sauvage}, M., {Sidher}, S., {Spencer},
  L., {Swinyard}, B., {Ward-Thompson}, D., \& {White}, G.~J. 2010, A\&A, 518,
  L99

\bibitem[{{Andre} {et~al.}(2000){Andre}, {Ward-Thompson}, \&
  {Barsony}}]{Andre00}
{Andre}, P., {Ward-Thompson}, D., \& {Barsony}, M. 2000, Protostars and Planets
  IV, 59

\bibitem[{{Andre} {et~al.}(1996){Andre}, {Ward-Thompson}, \& {Motte}}]{Andre96}
{Andre}, P., {Ward-Thompson}, D., \& {Motte}, F. 1996, A\&A, 314, 625

\bibitem[{Azimlu(2010)}]{Azimlu10}
Azimlu, M. 2010, PhD thesis, University of Waterloo

\bibitem[{{Azimlu} \& {Fich}(2011)}]{Azimlu11}
{Azimlu}, M., \& {Fich}, M. 2011, AJ, 141, 123

\bibitem[{{Beckwith} \& {Sargent}(1991)}]{Beck91}
{Beckwith}, S.~V.~W., \& {Sargent}, A.~I. 1991, ApJ, 381, 250

\bibitem[{{Beuther} {et~al.}(2002){Beuther}, {Schilke}, {Menten}, {Motte},
  {Sridharan}, \& {Wyrowski}}]{Beuther02}
{Beuther}, H., {Schilke}, P., {Menten}, K.~M., {Motte}, F., {Sridharan}, T.~K.,
  \& {Wyrowski}, F. 2002, ApJ, 566, 945

\bibitem[{{Bolatto} {et~al.}(2013){Bolatto}, {Wolfire}, \& {Leroy}}]{Bolatto13}
{Bolatto}, A.~D., {Wolfire}, M., \& {Leroy}, A.~K. 2013, ArXiv
  e-prints:1301.3498

\bibitem[{{Chapin} {et~al.}(2013){Chapin}, {Berry}, {Gibb}, {Jenness}, {Scott},
  {Tilanus}, {Economou}, \& {Holland}}]{Chapin13}
{Chapin}, E.~L., {Berry}, D.~S., {Gibb}, A.~G., {Jenness}, T., {Scott}, D.,
  {Tilanus}, R.~P.~J., {Economou}, F., \& {Holland}, W.~S. 2013, MNRAS, 430,
  2545

\bibitem[{{Chavarr{\'{\i}}a} {et~al.}(2008){Chavarr{\'{\i}}a}, {Allen}, {Hora},
  {Brunt}, \& {Fazio}}]{Fazio08}
{Chavarr{\'{\i}}a}, L.~A., {Allen}, L.~E., {Hora}, J.~L., {Brunt}, C.~M., \&
  {Fazio}, G.~G. 2008, ApJ, 682, 445

\bibitem[{{Chokshi} {et~al.}(1993){Chokshi}, {Tielens}, \&
  {Hollenbach}}]{Chokshi93}
{Chokshi}, A., {Tielens}, A.~G.~G.~M., \& {Hollenbach}, D. 1993, ApJ, 407, 806

\bibitem[{{Crampton} {et~al.}(1978){Crampton}, {Georgelin}, \&
  {Georgelin}}]{Crampton78}
{Crampton}, D., {Georgelin}, Y.~M., \& {Georgelin}, Y.~P. 1978, A\&A, 66, 1

\bibitem[{{Dempsey} {et~al.}(2013){Dempsey}, {Friberg}, {Jenness}, {Tilanus},
  {Thomas}, {Holland}, {Bintley}, {Berry}, {Chapin}, {Chrysostomou}, {Davis},
  {Gibb}, {Parsons}, \& {Robson}}]{Dempsey13}
{Dempsey}, J.~T., {Friberg}, P., {Jenness}, T., {Tilanus}, R.~P.~J., {Thomas},
  H.~S., {Holland}, W.~S., {Bintley}, D., {Berry}, D.~S., {Chapin}, E.~L.,
  {Chrysostomou}, A., {Davis}, G.~R., {Gibb}, A.~G., {Parsons}, H., \&
  {Robson}, E.~I. 2013, MNRAS, 430, 2534

\bibitem[{{D{\'e}sert} \& et~al.(2008)}]{Desert08}
{D{\'e}sert}, F., \& et~al. 2008, A\& A, 481, 411

\bibitem[{{Dickman}(1978)}]{Dickman78}
{Dickman}, R.~L. 1978, ApJS, 37, 407

\bibitem[{{Drabek} {et~al.}(2012){Drabek}, {Hatchell}, {Friberg}, {Richer},
  {Graves}, {Buckle}, {Nutter}, {Johnstone}, \& {Di Francesco}}]{Drabek12}
{Drabek}, E., {Hatchell}, J., {Friberg}, P., {Richer}, J., {Graves}, S.,
  {Buckle}, J.~V., {Nutter}, D., {Johnstone}, D., \& {Di Francesco}, J. 2012,
  MNRAS, 426, 23

\bibitem[{{Draine}(2003)}]{Draine03}
{Draine}, B.~T. 2003, ARA\& A, 41, 241

\bibitem[{{Draine} \& {Lee}(1984)}]{draine_lee84}
{Draine}, B.~T., \& {Lee}, H.~M. 1984, ApJ, 285, 89

\bibitem[{{Dupac} {et~al.}(2003){Dupac}, {Bernard}, {Boudet}, {Giard},
  {Lamarre}, {M{\'e}ny}, {Pajot}, {Ristorcelli}, {Serra}, {Stepnik}, \&
  {Torre}}]{Dupac03}
{Dupac}, X., {Bernard}, J., {Boudet}, N., {Giard}, M., {Lamarre}, J.,
  {M{\'e}ny}, C., {Pajot}, F., {Ristorcelli}, I., {Serra}, G., {Stepnik}, B.,
  \& {Torre}, J. 2003, A\& A, 404, L11

\bibitem[{{Evans} {et~al.}(2001){Evans}, {Rawlings}, {Shirley}, \&
  {Mundy}}]{Evans01}
{Evans}, II, N.~J., {Rawlings}, J.~M.~C., {Shirley}, Y.~L., \& {Mundy}, L.~G.
  2001, ApJ, 557, 193

\bibitem[{{Galli} {et~al.}(2002){Galli}, {Walmsley}, \& {Gon{\c
  c}alves}}]{Galli02}
{Galli}, D., {Walmsley}, M., \& {Gon{\c c}alves}, J. 2002, A\&A, 394, 275

\bibitem[{{Giannini} {et~al.}(2012){Giannini}, {Elia}, {Lorenzetti},
  {Molinari}, {Motte}, {Schisano}, {Pezzuto}, {Pestalozzi}, {di Giorgio},
  {Andr{\'e}}, {Hill}, {Benedettini}, {Bontemps}, {di Francesco}, {Fallscheer},
  {Hennemann}, {Kirk}, {Minier}, {Nguyen Luong}, {Polychroni}, {Rygl},
  {Saraceno}, {Schneider}, {Spinoglio}, {Testi}, {Ward-Thompson}, \&
  {White}}]{Giannini12}
{Giannini}, T., {Elia}, D., {Lorenzetti}, D., {Molinari}, S., {Motte}, F.,
  {Schisano}, E., {Pezzuto}, S., {Pestalozzi}, M., {di Giorgio}, A.~M.,
  {Andr{\'e}}, P., {Hill}, T., {Benedettini}, M., {Bontemps}, S., {di
  Francesco}, J., {Fallscheer}, C., {Hennemann}, M., {Kirk}, J., {Minier}, V.,
  {Nguyen Luong}, Q., {Polychroni}, D., {Rygl}, K.~L.~J., {Saraceno}, P.,
  {Schneider}, N., {Spinoglio}, L., {Testi}, L., {Ward-Thompson}, D., \&
  {White}, G.~J. 2012, A\&A, 539, A156

\bibitem[{{Goldsmith}(2001)}]{Goldsmith01}
{Goldsmith}, P.~F. 2001, ApJ, 557, 736

\bibitem[{{Gordon}(1995)}]{Gordon95}
{Gordon}, M.~A. 1995, A\&A, 301, 853

\bibitem[{{Hatchell} {et~al.}(2013){Hatchell}, {Wilson}, {Drabek}, {Curtis},
  {Richer}, {Nutter}, {Di Francesco}, {Ward-Thompson}, \& {JCMT GBS
  Consortium}}]{Hatchell13}
{Hatchell}, J., {Wilson}, T., {Drabek}, E., {Curtis}, E., {Richer}, J.,
  {Nutter}, D., {Di Francesco}, J., {Ward-Thompson}, D., \& {JCMT GBS
  Consortium}. 2013, MNRAS, 429, L10

\bibitem[{{Heyer} {et~al.}(1989){Heyer}, {Snell}, {Morgan}, \&
  {Schloerb}}]{Heyer89}
{Heyer}, M.~H., {Snell}, R.~L., {Morgan}, J., \& {Schloerb}, F.~P. 1989, ApJ,
  346, 220

\bibitem[{{Holland} {et~al.}(2013){Holland}, {Bintley}, {Chapin},
  {Chrysostomou}, {Davis}, {Dempsey}, {Duncan}, {Fich}, {Friberg}, {Halpern},
  {Irwin}, {Jenness}, {Kelly}, {MacIntosh}, {Robson}, {Scott}, {Ade},
  {Atad-Ettedgui}, {Berry}, {Craig}, {Gao}, {Gibb}, {Hilton}, {Hollister},
  {Kycia}, {Lunney}, {McGregor}, {Montgomery}, {Parkes}, {Tilanus}, {Ullom},
  {Walther}, {Walton}, {Woodcraft}, {Amiri}, {Atkinson}, {Burger}, {Chuter},
  {Coulson}, {Doriese}, {Dunare}, {Economou}, {Niemack}, {Parsons},
  {Reintsema}, {Sibthorpe}, {Smail}, {Sudiwala}, \& {Thomas}}]{Holland13}
{Holland}, W.~S., {Bintley}, D., {Chapin}, E.~L., {Chrysostomou}, A., {Davis},
  G.~R., {Dempsey}, J.~T., {Duncan}, W.~D., {Fich}, M., {Friberg}, P.,
  {Halpern}, M., {Irwin}, K.~D., {Jenness}, T., {Kelly}, B.~D., {MacIntosh},
  M.~J., {Robson}, E.~I., {Scott}, D., {Ade}, P.~A.~R., {Atad-Ettedgui}, E.,
  {Berry}, D.~S., {Craig}, S.~C., {Gao}, X., {Gibb}, A.~G., {Hilton}, G.~C.,
  {Hollister}, M.~I., {Kycia}, J.~B., {Lunney}, D.~W., {McGregor}, H.,
  {Montgomery}, D., {Parkes}, W., {Tilanus}, R.~P.~J., {Ullom}, J.~N.,
  {Walther}, C.~A., {Walton}, A.~J., {Woodcraft}, A.~L., {Amiri}, M.,
  {Atkinson}, D., {Burger}, B., {Chuter}, T., {Coulson}, I.~M., {Doriese},
  W.~B., {Dunare}, C., {Economou}, F., {Niemack}, M.~D., {Parsons}, H.~A.~L.,
  {Reintsema}, C.~D., {Sibthorpe}, B., {Smail}, I., {Sudiwala}, R., \&
  {Thomas}, H.~S. 2013, MNRAS, 430, 2513

\bibitem[{{Jenness} {et~al.}(2011){Jenness}, {Berry}, {Chapin}, {Economou},
  {Gibb}, \& {Scott}}]{JennessT2011}
{Jenness}, T., {Berry}, D., {Chapin}, E., {Economou}, F., {Gibb}, A., \&
  {Scott}, D. 2011, in Astronomical Society of the Pacific Conference Series,
  Vol. 442, Astronomical Data Analysis Software and Systems XX, ed.
  {I.~N.~Evans, A.~Accomazzi, D.~J.~Mink, \& A.~H.~Rots}, 281

\bibitem[{{Jenness} {et~al.}(2008){Jenness}, {Cavanagh}, {Economou}, \&
  {Berry}}]{JennessT08}
{Jenness}, T., {Cavanagh}, B., {Economou}, F., \& {Berry}, D.~S. 2008, in
  Astronomical Society of the Pacific Conference Series, Vol. 394, Astronomical
  Data Analysis Software and Systems XVII, ed. R.~W. {Argyle}, P.~S.
  {Bunclark}, \& J.~R. {Lewis}, 565

\bibitem[{{Jenness} \& {Economou}(1999)}]{oracdr}
{Jenness}, T., \& {Economou}, F. 1999, in Astronomical Society of the Pacific
  Conference Series, Vol. 172, Astronomical Data Analysis Software and Systems
  VIII, ed. D.~M. {Mehringer}, R.~L. {Plante}, \& D.~A. {Roberts}, 171

\bibitem[{{Johnstone} \& {Bally}(2006)}]{Johnstone06b}
{Johnstone}, D., \& {Bally}, J. 2006, ApJ, 653, 383

\bibitem[{{Johnstone} {et~al.}(2001){Johnstone}, {Fich}, {Mitchell}, \&
  {Moriarty-Schieven}}]{Johnstone01}
{Johnstone}, D., {Fich}, M., {Mitchell}, G.~F., \& {Moriarty-Schieven}, G.
  2001, ApJ, 559, 307

\bibitem[{{Johnstone} {et~al.}(2006){Johnstone}, {Matthews}, \&
  {Mitchell}}]{Johnstone06a}
{Johnstone}, D., {Matthews}, H., \& {Mitchell}, G.~F. 2006, ApJ, 639, 259

\bibitem[{{J{\o}rgensen} {et~al.}(2006){J{\o}rgensen}, {Johnstone}, {van
  Dishoeck}, \& {Doty}}]{Jorgensen06}
{J{\o}rgensen}, J.~K., {Johnstone}, D., {van Dishoeck}, E.~F., \& {Doty}, S.~D.
  2006, A\&A, 449, 609

\bibitem[{{J{\o}rgensen} {et~al.}(2002){J{\o}rgensen}, {Sch{\"o}ier}, \& {van
  Dishoeck}}]{Jorgensen02}
{J{\o}rgensen}, J.~K., {Sch{\"o}ier}, F.~L., \& {van Dishoeck}, E.~F. 2002,
  A\&A, 389, 908

\bibitem[{{Juvela} {et~al.}(2013){Juvela}, {Montillaud}, {Ysard}, \&
  {Lunttila}}]{Juvela13}
{Juvela}, M., {Montillaud}, J., {Ysard}, N., \& {Lunttila}, T. 2013, A\&A, 556,
  A63

\bibitem[{{Juvela} {et~al.}(2011){Juvela}, {Ristorcelli}, {Pelkonen},
  {Marshall}, {Montier}, {Bernard}, {Paladini}, {Lunttila}, {Abergel},
  {Andr{\'e}}, {Dickinson}, {Dupac}, {Malinen}, {Martin}, {McGehee}, {Pagani},
  {Ysard}, \& {Zavagno}}]{Juvela11}
{Juvela}, M., {Ristorcelli}, I., {Pelkonen}, V., {Marshall}, D.~J., {Montier},
  L.~A., {Bernard}, J., {Paladini}, R., {Lunttila}, T., {Abergel}, A.,
  {Andr{\'e}}, P., {Dickinson}, C., {Dupac}, X., {Malinen}, J., {Martin}, P.,
  {McGehee}, P., {Pagani}, L., {Ysard}, N., \& {Zavagno}, A. 2011, A\&A, 527,
  A111

\bibitem[{{Kackley} {et~al.}(2010){Kackley}, {Scott}, {Chapin}, \&
  {Friberg}}]{Kackley10}
{Kackley}, R., {Scott}, D., {Chapin}, E., \& {Friberg}, P. 2010, in Society of
  Photo-Optical Instrumentation Engineers (SPIE) Conference Series, Vol. 7740,
  Society of Photo-Optical Instrumentation Engineers (SPIE) Conference Series

\bibitem[{{Kauffmann} \& {Bertoldi}(2004)}]{Kauf04}
{Kauffmann}, J., \& {Bertoldi}, F. 2004, ArXiv eprints:astro-ph/0402021

\bibitem[{{Kelly} {et~al.}(2012){Kelly}, {Shetty}, {Stutz}, {Kauffmann},
  {Goodman}, \& {Launhardt}}]{Kelly12}
{Kelly}, B.~C., {Shetty}, R., {Stutz}, A.~M., {Kauffmann}, J., {Goodman},
  A.~A., \& {Launhardt}, R. 2012, ApJ, 752, 55

\bibitem[{{Kramer} {et~al.}(2003){Kramer}, {Richer}, {Mookerjea}, {Alves}, \&
  {Lada}}]{Kramer02}
{Kramer}, C., {Richer}, J., {Mookerjea}, B., {Alves}, J., \& {Lada}, C. 2003,
  A\&A, 399, 1073

\bibitem[{{Kuan} {et~al.}(1996){Kuan}, {Mehringer}, \& {Snyder}}]{Kuan96}
{Kuan}, Y.-J., {Mehringer}, D.~M., \& {Snyder}, L.~E. 1996, ApJ, 459, 619

\bibitem[{{Kwon} {et~al.}(2009){Kwon}, {Looney}, {Mundy}, {Chiang}, \&
  {Kemball}}]{Kwon09}
{Kwon}, W., {Looney}, L.~W., {Mundy}, L.~G., {Chiang}, H.-F., \& {Kemball},
  A.~J. 2009, ApJ, 696, 841

\bibitem[{{Lada}(1987)}]{Lada87}
{Lada}, C.~J. 1987, in IAU Symposium, Vol. 115, Star Forming Regions, ed.
  M.~{Peimbert} \& J.~{Jugaku}, 1--17

\bibitem[{{Launhardt} {et~al.}(2013){Launhardt}, {Stutz}, {Schmiedeke},
  {Henning}, {Krause}, {Balog}, {Beuther}, {Birkmann}, {Hennemann},
  {Kainulainen}, {Khanzadyan}, {Linz}, {Lippok}, {Nielbock}, {Pitann}, {Ragan},
  {Risacher}, {Schmalzl}, {Shirley}, {Stecklum}, {Steinacker}, \&
  {Tackenberg}}]{Laun13}
{Launhardt}, R., {Stutz}, A.~M., {Schmiedeke}, A., {Henning}, T., {Krause}, O.,
  {Balog}, Z., {Beuther}, H., {Birkmann}, S., {Hennemann}, M., {Kainulainen},
  J., {Khanzadyan}, T., {Linz}, H., {Lippok}, N., {Nielbock}, M., {Pitann}, J.,
  {Ragan}, S., {Risacher}, C., {Schmalzl}, M., {Shirley}, Y.~L., {Stecklum},
  B., {Steinacker}, J., \& {Tackenberg}, J. 2013, A\&A, 551, A98

\bibitem[{{Leroy} {et~al.}(2009){Leroy}, {Bolatto}, {Bot}, {Engelbracht},
  {Gordon}, {Israel}, {Rubio}, {Sandstrom}, \& {Stanimirovi{\'c}}}]{Leory09}
{Leroy}, A.~K., {Bolatto}, A., {Bot}, C., {Engelbracht}, C.~W., {Gordon}, K.,
  {Israel}, F.~P., {Rubio}, M., {Sandstrom}, K., \& {Stanimirovi{\'c}}, S.
  2009, ApJ, 702, 352

\bibitem[{{Lis} {et~al.}(1998){Lis}, {Serabyn}, {Keene}, {Dowell}, {Benford},
  {Phillips}, {Hunter}, \& {Wang}}]{Lis98}
{Lis}, D.~C., {Serabyn}, E., {Keene}, J., {Dowell}, C.~D., {Benford}, D.~J.,
  {Phillips}, T.~G., {Hunter}, T.~R., \& {Wang}, N. 1998, ApJ, 509, 299

\bibitem[{{Mathis}(1990)}]{Mathis90}
{Mathis}, J.~S. 1990, ARA\&A, 28, 37

\bibitem[{{Matzner} \& {McKee}(2000)}]{Matzner00}
{Matzner}, C.~D., \& {McKee}, C.~F. 2000, ApJ, 545, 364

\bibitem[{{Miyake} \& {Nakagawa}(1993)}]{Miyake93}
{Miyake}, K., \& {Nakagawa}, Y. 1993, Icarus, 106, 20

\bibitem[{{Moffat} {et~al.}(1979){Moffat}, {Jackson}, \&
  {Fitzgerald}}]{Moffat79}
{Moffat}, A.~F.~J., {Jackson}, P.~D., \& {Fitzgerald}, M.~P. 1979, A\&AS, 38,
  197

\bibitem[{{Mookerjea} {et~al.}(2000){Mookerjea}, {Ghosh}, {Rengarajan},
  {Tandon}, \& {Verma}}]{Mookerjea00}
{Mookerjea}, B., {Ghosh}, S.~K., {Rengarajan}, T.~N., {Tandon}, S.~N., \&
  {Verma}, R.~P. 2000, AJ, 120, 1954

\bibitem[{{Motte} {et~al.}(2010){Motte}, {Zavagno}, {Bontemps}, {Schneider},
  {Hennemann}, {di Francesco}, {Andr{\'e}}, {Saraceno}, {Griffin}, {Marston},
  {Ward-Thompson}, {White}, {Minier}, {Men'shchikov}, {Hill}, {Abergel},
  {Anderson}, {Aussel}, {Balog}, {Baluteau}, {Bernard}, {Cox}, {Csengeri},
  {Deharveng}, {Didelon}, {di Giorgio}, {Hargrave}, {Huang}, {Kirk}, {Leeks},
  {Li}, {Martin}, {Molinari}, {Nguyen-Luong}, {Olofsson}, {Persi}, {Peretto},
  {Pezzuto}, {Roussel}, {Russeil}, {Sadavoy}, {Sauvage}, {Sibthorpe},
  {Spinoglio}, {Testi}, {Teyssier}, {Vavrek}, {Wilson}, \&
  {Woodcraft}}]{Motte10}
{Motte}, F., {Zavagno}, A., {Bontemps}, S., {Schneider}, N., {Hennemann}, M.,
  {di Francesco}, J., {Andr{\'e}}, P., {Saraceno}, P., {Griffin}, M.,
  {Marston}, A., {Ward-Thompson}, D., {White}, G., {Minier}, V.,
  {Men'shchikov}, A., {Hill}, T., {Abergel}, A., {Anderson}, L.~D., {Aussel},
  H., {Balog}, Z., {Baluteau}, J.-P., {Bernard}, J.-P., {Cox}, P., {Csengeri},
  T., {Deharveng}, L., {Didelon}, P., {di Giorgio}, A.-M., {Hargrave}, P.,
  {Huang}, M., {Kirk}, J., {Leeks}, S., {Li}, J.~Z., {Martin}, P., {Molinari},
  S., {Nguyen-Luong}, Q., {Olofsson}, G., {Persi}, P., {Peretto}, N.,
  {Pezzuto}, S., {Roussel}, H., {Russeil}, D., {Sadavoy}, S., {Sauvage}, M.,
  {Sibthorpe}, B., {Spinoglio}, L., {Testi}, L., {Teyssier}, D., {Vavrek}, R.,
  {Wilson}, C.~D., \& {Woodcraft}, A. 2010, A\&A, 518, L77

\bibitem[{{Mueller} {et~al.}(2002){Mueller}, {Shirley}, {Evans}, \&
  {Jacobson}}]{Mueller02}
{Mueller}, K.~E., {Shirley}, Y.~L., {Evans}, II, N.~J., \& {Jacobson}, H.~R.
  2002, APJS, 143, 469

\bibitem[{{Myers} {et~al.}(1986){Myers}, {Dame}, {Thaddeus}, {Cohen},
  {Silverberg}, {Dwek}, \& {Hauser}}]{Myers86}
{Myers}, P.~C., {Dame}, T.~M., {Thaddeus}, P., {Cohen}, R.~S., {Silverberg},
  R.~F., {Dwek}, E., \& {Hauser}, M.~G. 1986, ApJ, 301, 398

\bibitem[{{Papadopoulos} \& {Allen}(2000)}]{Papa00}
{Papadopoulos}, P.~P., \& {Allen}, M.~L. 2000, ApJ, 537, 631

\bibitem[{{Paradis} {et~al.}(2010){Paradis}, {Veneziani}, {Noriega-Crespo},
  {Paladini}, {Piacentini}, {Bernard}, {de Bernardis}, {Calzoletti},
  {Faustini}, {Martin}, {Masi}, {Montier}, {Natoli}, {Ristorcelli}, {Thompson},
  {Traficante}, \& {Molinari}}]{Paradis10}
{Paradis}, D., {Veneziani}, M., {Noriega-Crespo}, A., {Paladini}, R.,
  {Piacentini}, F., {Bernard}, J.~P., {de Bernardis}, P., {Calzoletti}, L.,
  {Faustini}, F., {Martin}, P., {Masi}, S., {Montier}, L., {Natoli}, P.,
  {Ristorcelli}, I., {Thompson}, M.~A., {Traficante}, A., \& {Molinari}, S.
  2010, A\&A, 520, L8

\bibitem[{{Pascale} {et~al.}(2008){Pascale}, {Ade}, {Bock}, {Chapin}, {Chung},
  {Devlin}, {Dicker}, {Griffin}, {Gundersen}, {Halpern}, {Hargrave}, {Hughes},
  {Klein}, {MacTavish}, {Marsden}, {Martin}, {Martin}, {Mauskopf},
  {Netterfield}, {Olmi}, {Patanchon}, {Rex}, {Scott}, {Semisch}, {Thomas},
  {Truch}, {Tucker}, {Tucker}, {Viero}, \& {Wiebe}}]{Pascale08}
{Pascale}, E., {Ade}, P.~A.~R., {Bock}, J.~J., {Chapin}, E.~L., {Chung}, J.,
  {Devlin}, M.~J., {Dicker}, S., {Griffin}, M., {Gundersen}, J.~O., {Halpern},
  M., {Hargrave}, P.~C., {Hughes}, D.~H., {Klein}, J., {MacTavish}, C.~J.,
  {Marsden}, G., {Martin}, P.~G., {Martin}, T.~G., {Mauskopf}, P.,
  {Netterfield}, C.~B., {Olmi}, L., {Patanchon}, G., {Rex}, M., {Scott}, D.,
  {Semisch}, C., {Thomas}, N., {Truch}, M.~D.~P., {Tucker}, C., {Tucker},
  G.~S., {Viero}, M.~P., \& {Wiebe}, D.~V. 2008, ApJ, 681, 400

\bibitem[{{Pilbratt} {et~al.}(2010){Pilbratt}, {Riedinger}, {Passvogel},
  {Crone}, {Doyle}, {Gageur}, {Heras}, {Jewell}, {Metcalfe}, {Ott}, \&
  {Schmidt}}]{Pilbratt10}
{Pilbratt}, G.~L., {Riedinger}, J.~R., {Passvogel}, T., {Crone}, G., {Doyle},
  D., {Gageur}, U., {Heras}, A.~M., {Jewell}, C., {Metcalfe}, L., {Ott}, S., \&
  {Schmidt}, M. 2010, A\&A, 518, L1

\bibitem[{{Pirogov}(2009)}]{Pirogov09}
{Pirogov}, L.~E. 2009, Astronomy Reports, 53, 1127

\bibitem[{{Pismis} \& {Hasse}(1976)}]{Pismis76}
{Pismis}, P., \& {Hasse}, I. 1976, AP\&SS, 45, 79

\bibitem[{{Planck Collaboration} {et~al.}(2011){Planck Collaboration}, {Ade},
  {Aghanim}, {Arnaud}, {Ashdown}, {Aumont}, {Baccigalupi}, {Balbi}, {Banday},
  {Barreiro}, \& et~al.}]{PlanckC11}
{Planck Collaboration}, {Ade}, P.~A.~R., {Aghanim}, N., {Arnaud}, M.,
  {Ashdown}, M., {Aumont}, J., {Baccigalupi}, C., {Balbi}, A., {Banday}, A.~J.,
  {Barreiro}, R.~B., \& et~al. 2011, A\&A, 536, A22

\bibitem[{{Ricci} {et~al.}(2010{\natexlab{a}}){Ricci}, {Testi}, {Natta}, \&
  {Brooks}}]{Ricci10b}
{Ricci}, L., {Testi}, L., {Natta}, A., \& {Brooks}, K.~J. 2010{\natexlab{a}},
  A\&A, 521, A66

\bibitem[{{Ricci} {et~al.}(2010{\natexlab{b}}){Ricci}, {Testi}, {Natta},
  {Neri}, {Cabrit}, \& {Herczeg}}]{Ricci10a}
{Ricci}, L., {Testi}, L., {Natta}, A., {Neri}, R., {Cabrit}, S., \& {Herczeg},
  G.~J. 2010{\natexlab{b}}, A\&A, 512, A15

\bibitem[{{Ricci} {et~al.}(2012){Ricci}, {Testi}, {Natta}, {Scholz}, \& {de
  Gregorio-Monsalvo}}]{Ricci12}
{Ricci}, L., {Testi}, L., {Natta}, A., {Scholz}, A., \& {de Gregorio-Monsalvo},
  I. 2012, ApJL, 761, L20

\bibitem[{{Roy} {et~al.}(2011){Roy}, {Ade}, {Bock}, {Chapin}, {Devlin},
  {Dicker}, {France}, {Gibb}, {Griffin}, {Gundersen}, {Halpern}, {Hargrave},
  {Hughes}, {Klein}, {Marsden}, {Martin}, {Mauskopf}, {Morales Ortiz},
  {Netterfield}, {Noriega-Crespo}, {Olmi}, {Patanchon}, {Rex}, {Scott},
  {Semisch}, {Truch}, {Tucker}, {Tucker}, {Viero}, \& {Wiebe}}]{Roy11a}
{Roy}, A., {Ade}, P.~A.~R., {Bock}, J.~J., {Chapin}, E.~L., {Devlin}, M.~J.,
  {Dicker}, S.~R., {France}, K., {Gibb}, A.~G., {Griffin}, M., {Gundersen},
  J.~O., {Halpern}, M., {Hargrave}, P.~C., {Hughes}, D.~H., {Klein}, J.,
  {Marsden}, G., {Martin}, P.~G., {Mauskopf}, P., {Morales Ortiz}, J.~L.,
  {Netterfield}, C.~B., {Noriega-Crespo}, A., {Olmi}, L., {Patanchon}, G.,
  {Rex}, M., {Scott}, D., {Semisch}, C., {Truch}, M.~D.~P., {Tucker}, C.,
  {Tucker}, G.~S., {Viero}, M.~P., \& {Wiebe}, D.~V. 2011, ApJ, 727, 114

\bibitem[{{Russeil} {et~al.}(2007){Russeil}, {Adami}, \& {Georgelin}}]{Rus07}
{Russeil}, D., {Adami}, C., \& {Georgelin}, Y.~M. 2007, A\& A, 470, 161

\bibitem[{{Russeil} {et~al.}(1995){Russeil}, {Georgelin}, {Georgelin}, {Le
  Coarer}, \& {Marcelin}}]{Russeil95}
{Russeil}, D., {Georgelin}, Y.~M., {Georgelin}, Y.~P., {Le Coarer}, E., \&
  {Marcelin}, M. 1995, A\&AS, 114, 557

\bibitem[{{Sadavoy} {et~al.}(2013){Sadavoy}, {Di Francesco}, {Johnstone},
  {Currie}, {Drabek}, {Hatchell}, {Nutter}, {Andr{\'e}}, {Arzoumanian},
  {Benedettini}, {Bernard}, {Duarte-Cabral}, {Fallscheer}, {Friesen},
  {Greaves}, {Hennemann}, {Hill}, {Jenness}, {K{\"o}nyves}, {Matthews},
  {Mottram}, {Pezzuto}, {Roy}, {Rygl}, {Schneider-Bontemps}, {Spinoglio},
  {Testi}, {Tothill}, {Ward-Thompson}, {White}, \& {JCMT and Herschel Gould
  Belt Survey teams}}]{Sadavoy13}
{Sadavoy}, S.~I., {Di Francesco}, J., {Johnstone}, D., {Currie}, M.~J.,
  {Drabek}, E., {Hatchell}, J., {Nutter}, D., {Andr{\'e}}, P., {Arzoumanian},
  D., {Benedettini}, M., {Bernard}, J.-P., {Duarte-Cabral}, A., {Fallscheer},
  C., {Friesen}, R., {Greaves}, J., {Hennemann}, M., {Hill}, T., {Jenness}, T.,
  {K{\"o}nyves}, V., {Matthews}, B., {Mottram}, J.~C., {Pezzuto}, S., {Roy},
  A., {Rygl}, K., {Schneider-Bontemps}, N., {Spinoglio}, L., {Testi}, L.,
  {Tothill}, N., {Ward-Thompson}, D., {White}, G., \& {JCMT and Herschel Gould
  Belt Survey teams}. 2013, ApJ, 767, 126

\bibitem[{{Schnee} {et~al.}(2010){Schnee}, {Enoch}, {Noriega-Crespo}, {Sayers},
  {Terebey}, {Caselli}, {Foster}, {Goodman}, {Kauffmann}, {Padgett}, {Rebull},
  {Sargent}, \& {Shetty}}]{Schnee10}
{Schnee}, S., {Enoch}, M., {Noriega-Crespo}, A., {Sayers}, J., {Terebey}, S.,
  {Caselli}, P., {Foster}, J., {Goodman}, A., {Kauffmann}, J., {Padgett}, D.,
  {Rebull}, L., {Sargent}, A., \& {Shetty}, R. 2010, ApJ, 708, 127

\bibitem[{{Schnee} \& {Goodman}(2005)}]{Schnee05}
{Schnee}, S., \& {Goodman}, A. 2005, ApJ, 624, 254

\bibitem[{{Seaquist} {et~al.}(2004){Seaquist}, {Yao}, {Dunne}, \&
  {Cameron}}]{Seaquist04}
{Seaquist}, E., {Yao}, L., {Dunne}, L., \& {Cameron}, H. 2004, MNRAS, 349, 1428

\bibitem[{{Shetty} {et~al.}(2009{\natexlab{a}}){Shetty}, {Kauffmann}, {Schnee},
  \& {Goodman}}]{Shetty09a}
{Shetty}, R., {Kauffmann}, J., {Schnee}, S., \& {Goodman}, A.~A.
  2009{\natexlab{a}}, ApJ, 696, 676

\bibitem[{{Shetty} {et~al.}(2009{\natexlab{b}}){Shetty}, {Kauffmann}, {Schnee},
  {Goodman}, \& {Ercolano}}]{Shetty09b}
{Shetty}, R., {Kauffmann}, J., {Schnee}, S., {Goodman}, A.~A., \& {Ercolano},
  B. 2009{\natexlab{b}}, ApJ, 696, 2234

\bibitem[{{Shirley} {et~al.}(2000){Shirley}, {Evans}, {Rawlings}, \&
  {Gregersen}}]{Shirley00}
{Shirley}, Y.~L., {Evans}, II, N.~J., {Rawlings}, J.~M.~C., \& {Gregersen},
  E.~M. 2000, ApJS, 131, 249

\bibitem[{{Shirley} {et~al.}(2011){Shirley}, {Huard}, {Pontoppidan}, {Wilner},
  {Stutz}, {Bieging}, \& {Evans}}]{Shirley11}
{Shirley}, Y.~L., {Huard}, T.~L., {Pontoppidan}, K.~M., {Wilner}, D.~J.,
  {Stutz}, A.~M., {Bieging}, J.~H., \& {Evans}, II, N.~J. 2011, ApJ, 728, 143

\bibitem[{{Shirley} {et~al.}(2005){Shirley}, {Nordhaus}, {Grcevich}, {Evans},
  {Rawlings}, \& {Tatematsu}}]{Shirley05}
{Shirley}, Y.~L., {Nordhaus}, M.~K., {Grcevich}, J.~M., {Evans}, II, N.~J.,
  {Rawlings}, J.~M.~C., \& {Tatematsu}, K. 2005, ApJ, 632, 982

\bibitem[{{Shu}(1977)}]{Shu77}
{Shu}, F.~H. 1977, ApJ, 214, 488

\bibitem[{Sreenilayam(2012)}]{Gopika12}
Sreenilayam, G. 2012, PhD thesis, University of Waterloo

\bibitem[{{Sridharan} {et~al.}(2005){Sridharan}, {Beuther}, {Saito},
  {Wyrowski}, \& {Schilke}}]{Sridharan05}
{Sridharan}, T.~K., {Beuther}, H., {Saito}, M., {Wyrowski}, F., \& {Schilke},
  P. 2005, ApJL, 634, L57

\bibitem[{{Stamatellos} {et~al.}(2007){Stamatellos}, {Whitworth}, \&
  {Ward-Thompson}}]{Stam07}
{Stamatellos}, D., {Whitworth}, A.~P., \& {Ward-Thompson}, D. 2007, MNRAS, 379,
  1390

\bibitem[{{Swift} \& {Welch}(2008)}]{Swift08}
{Swift}, J.~J., \& {Welch}, W.~J. 2008, ApJS, 174, 202

\bibitem[{{Tackenberg} {et~al.}(2012){Tackenberg}, {Beuther}, {Henning},
  {Schuller}, {Wienen}, {Motte}, {Wyrowski}, {Bontemps}, {Bronfman}, {Menten},
  {Testi}, \& {Lefloch}}]{Tackenberg12}
{Tackenberg}, J., {Beuther}, H., {Henning}, T., {Schuller}, F., {Wienen}, M.,
  {Motte}, F., {Wyrowski}, F., {Bontemps}, S., {Bronfman}, L., {Menten}, K.,
  {Testi}, L., \& {Lefloch}, B. 2012, A\&A, 540, A113

\bibitem[{{Tafalla} {et~al.}(2002){Tafalla}, {Myers}, {Caselli}, {Walmsley}, \&
  {Comito}}]{Tafalla02}
{Tafalla}, M., {Myers}, P.~C., {Caselli}, P., {Walmsley}, C.~M., \& {Comito},
  C. 2002, ApJ, 569, 815

\bibitem[{{Testi} {et~al.}(2003){Testi}, {Natta}, {Shepherd}, \&
  {Wilner}}]{Testi03}
{Testi}, L., {Natta}, A., {Shepherd}, D.~S., \& {Wilner}, D.~J. 2003, A\&A,
  403, 323

\bibitem[{{Ubach} {et~al.}(2012){Ubach}, {Maddison}, {Wright}, {Wilner},
  {Lommen}, \& {Koribalski}}]{Ubach12}
{Ubach}, C., {Maddison}, S.~T., {Wright}, C.~M., {Wilner}, D.~J., {Lommen},
  D.~J.~P., \& {Koribalski}, B. 2012, MNRAS, 425, 3137

\bibitem[{{van der Tak} {et~al.}(2000){van der Tak}, {van Dishoeck}, {Evans},
  \& {Blake}}]{Vtak00}
{van der Tak}, F.~F.~S., {van Dishoeck}, E.~F., {Evans}, II, N.~J., \& {Blake},
  G.~A. 2000, ApJ, 537, 283

\bibitem[{{van Dishoeck} \& {Black}(1988)}]{Dishoek88}
{van Dishoeck}, E.~F., \& {Black}, J.~H. 1988, ApJ, 334, 771

\bibitem[{{Ward} {et~al.}(2012){Ward}, {Wadsley}, {Sills}, \&
  {Petitclerc}}]{Ward12}
{Ward}, R.~L., {Wadsley}, J., {Sills}, A., \& {Petitclerc}, N. 2012, ApJ, 756,
  119

\bibitem[{{Ward-Thompson} {et~al.}(1999){Ward-Thompson}, {Motte}, \&
  {Andre}}]{Ward99}
{Ward-Thompson}, D., {Motte}, F., \& {Andre}, P. 1999, MNRAS, 305, 143

\bibitem[{{Ward-Thompson} {et~al.}(1994){Ward-Thompson}, {Scott}, {Hills}, \&
  {Andre}}]{Ward94}
{Ward-Thompson}, D., {Scott}, P.~F., {Hills}, R.~E., \& {Andre}, P. 1994,
  MNRAS, 268, 276

\bibitem[{{Whittet}(2003)}]{Whittet2003}
{Whittet}, D.~C.~B., ed. 2003, {Dust in the galactic environment}

\bibitem[{{Wilcock} {et~al.}(2012){Wilcock}, {Ward-Thompson}, {Kirk},
  {Stamatellos}, {Whitworth}, {Battersby}, {Elia}, {Fuller}, {DiGiorgio},
  {Griffin}, {Molinari}, {Martin}, {Mottram}, {Peretto}, {Pestalozzi},
  {Schisano}, {Smith}, \& {Thompson}}]{Wilcock12}
{Wilcock}, L.~A., {Ward-Thompson}, D., {Kirk}, J.~M., {Stamatellos}, D.,
  {Whitworth}, A., {Battersby}, C., {Elia}, D., {Fuller}, G.~A., {DiGiorgio},
  A., {Griffin}, M.~J., {Molinari}, S., {Martin}, P., {Mottram}, J.~C.,
  {Peretto}, N., {Pestalozzi}, M., {Schisano}, E., {Smith}, H.~A., \&
  {Thompson}, M.~A. 2012, MNRAS, 424, 716

\bibitem[{{Williams} {et~al.}(2005){Williams}, {Fuller}, \&
  {Sridharan}}]{williams05}
{Williams}, S.~J., {Fuller}, G.~A., \& {Sridharan}, T.~K. 2005, A\&A, 434, 257

\end{thebibliography}
\end{document}